%% file: main.tex
\documentclass[reprint,amsmath,amssymb,aps,showkeys,showpacs,nofootii]{revtex4-2}
\usepackage[caption=false]{subfig}
\usepackage{graphicx}
\usepackage{dcolumn}
\usepackage{bm}
\usepackage{float}
\usepackage{gensymb}
\usepackage{hyperref}
\usepackage{ragged2e}
\usepackage{lineno}
\usepackage{ulem}
\usepackage[table]{xcolor}
\newcommand\orcid[1]{\href{https://orcid.org/#1}{$\!$\includegraphics[scale=0.0045]{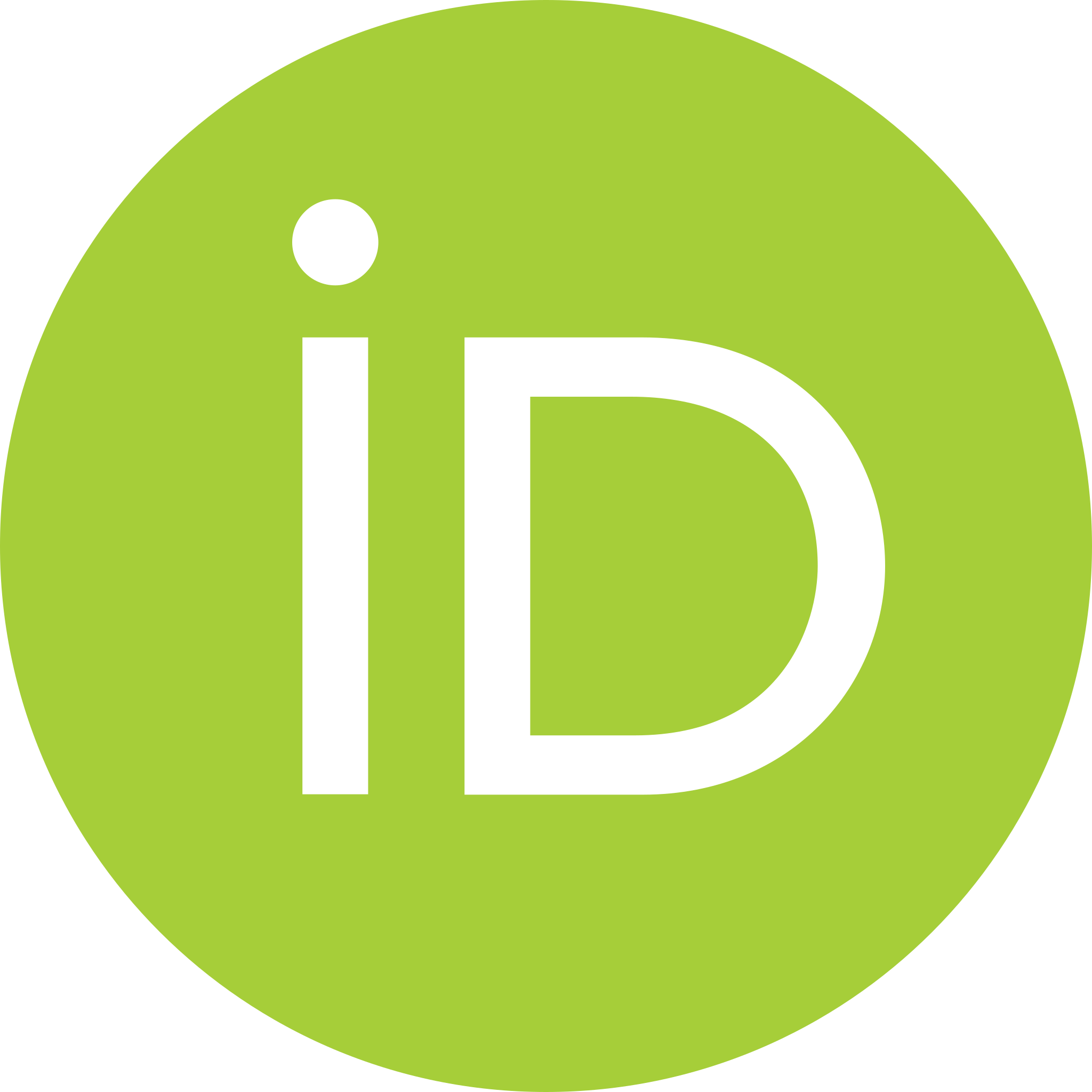} $\!\!$}}

\newcommand{\dif}{\textnormal{\slshape d}}
\newcommand{\Fermi}{\textit{Fermi }}

\usepackage{xcolor}

\begin{document}

\preprint{APS/123-QED}

\title{Searching for Axion-Like Particles from Core-Collapse Supernovae \\with \textit{Fermi} LAT's Low Energy Technique} 

\input{authors}
\date{\today}

\begin{abstract}
\input{abstract}
\end{abstract}

\pacs{xxxxxx}
\maketitle

%
\section{Introduction}\label{sec:intro}
\input{intro}

%
\section{ALP model}\label{sec:ALPmodel}
\input{ALP_model}

%
\section{Data Selection}\label{sec:dataselection}
\input{data_selection}

%
\section{Sensitivity Analysis}\label{sec:analysis}
\input{sensitivity}

%
\section{Search for an ALP signal within the selected GRB sample}\label{sec:GRBs}
\input{GRBs_analysis}

%
\section{Conclusions and Discussion}\label{sec:discussion and future work}
\input{conclusion}

%
\section*{Acknowledgments}\label{sec:ack}
\input{acknowledge}

%
\appendix
\input{appendix}


\bibliography{main}


\end{document}

%% file: authors.tex
\author{Milena Crnogor\v{c}evi\'{c} \orcid{0000-0002-7604-1779}}
\affiliation{Department of Astronomy, University of Maryland, College Park, MD 20742, USA}
\affiliation{Center for Research and Exploration in Space Science and Technology, NASA Goddard Space Flight Center, Greenbelt, MD 20771, USA}
\email{E-mail: mcrnogor@astro.umd.edu}

\author{Regina Caputo}
\affiliation{NASA Goddard Space Flight Center, Greenbelt, MD 20771, USA}

\author{Manuel Meyer}
\affiliation{Institute for Experimental Physics, University of Hamburg, Luruper Chaussee 149, 22671 Hamburg, Germany}

\author{Nicola Omodei}
\affiliation{W. W. Hansen Experimental Physics Laboratory, Kavli Institute for Particle Astrophysics and Cosmology, Department of Physics and SLAC National Accelerator Laboratory, Stanford University, Stanford, CA 94305, USA}

\author{Michael Gustafsson}
\affiliation{The Oskar Klein Centre for Cosmoparticle Physics, Department of Physics, Stockholm University, AlbaNova, SE-106 91 Stockholm, Sweden}

%% file: abstract.tex
Light axion-like particles (ALPs) are expected to be abundantly produced in core-collapse supernovae (CCSNe), resulting in a $\sim$10-second long burst of ALPs. These particles subsequently undergo conversion into gamma-rays in external magnetic fields to produce a long gamma-ray burst (GRB) with a  characteristic spectrum peaking in the 30--100-MeV energy range. At the same time, CCSNe are invoked as progenitors of {\it ordinary} long GRBs, rendering it relevant to conduct a comprehensive search for ALP spectral signatures using the observations of long GRB with the \textit{Fermi} Large Area Telescope (LAT). We perform a data-driven sensitivity analysis to determine CCSN distances for which a detection of an ALP signal is possible with the LAT's low-energy (LLE) technique which, in contrast to the standard LAT analysis, allows for a a larger effective area for energies down to 30~MeV. Assuming an ALP mass $m_a \lesssim 10^{-10}$~eV and ALP-photon coupling $g_{a\gamma} = 5.3\times 10^{-12}$ GeV$^{-1}$, values considered and deduced in ALP searches from SN1987A, we find that the distance limit ranges from $\sim\!0.5$ to $\sim\!10$~Mpc, depending on the sky location and the CCSN progenitor mass. Furthermore, we select a candidate sample of twenty-four GRBs and carry out a model comparison analysis in which we consider different GRB spectral models with and without an ALP signal component. We find that the inclusion of an ALP contribution does not result in any statistically significant improvement of the fits to the data. We discuss the statistical method used in our analysis and the underlying physical assumptions, the feasibility of setting upper limits on the ALP-photon coupling, and give an outlook on future telescopes in the context of ALP searches.

%% file: intro.tex

The axion-like particle (ALP), a generalized case of the QCD axion, belongs to the family of very weakly interacting subelectronvolt particles (WISPs) (see e.g.\ Ref.~\cite{Axions-review2,Jaeckel_2010,ringwald2014axions, Marsh_2016} and references within for a review of the QCD axion \cite{Peccei1, Peccei2,Axion-review3,Wilczek:1977pj}.)  The interaction of ALPs with photons can be described by the following Lagrangian,
\begin{equation}
\label{eq:L-alp}
    \mathcal{L}_{a\gamma} \supset  - \frac{1}{4} g_{a\gamma} \textbf{E}\cdot\textbf{B} \, a,
\end{equation}
where $g_{a\gamma}$ is the photon-ALP coupling, $\textbf{E}$ is the electric field, $\textbf{B}$ is the magnetic field, and $a$ represents the ALP field. When an external magnetic field $\textbf{B}$ is present, the two-photon coupling results in a photon-ALP conversion \cite{Raffelt:1987im}. Photon-ALP oscillations have been invoked to explain the excess of soft X-rays from the center of galaxy clusters \cite{Conlon:2013txa,Angus:2013sua,Powell:2014mda,Kraljic_2015}, the monochromatic 3.55-keV line in galaxy clusters \cite{Cicoli:2014bfa}, the low opacity of the Universe to
TeV photons \cite{Mirizzi:2009aj, opacity, Galanti:2015rda, Kohri:2017ljt, Zhou:2021usu}, anomalous stellar cooling \cite{Giannotti:2015kwo, Giannotti:2016hnk, Giannotti:2017hny, Dessert:2021bkv}, as well as the low-energy electronic recoil event excess in XENON1T \cite{XENON1T:2020tmw,DiLuzio:2020jjp,Gao:2020wer}. Furthermore, ALPs are considered one of the leading candidates for cold dark matter \cite{Choi:2020rgn, wispy, Preskill:1982cy, Abbott:1982af, Dine:1982ah}. The ALP parameter space has been explored using various experimental approaches, including light-shining-through-the-wall experiments \cite{Ehret:2010mh}, cavity experiments \cite{Duffy:2006aa}, as well as observations of  different astrophysical targets, such as Cepheid variable stars \cite{Friedland:2012hj}, star clusters \cite{Ayala:2014pea, Dessert:2020lil}, and galaxy clusters \cite{TheFermi-LAT:2016zue, Cheng:2020bhr, Marsh_2017, Malyshev:2018rsh, Reynolds:2019uqt}.
\par 
In this paper, we investigate the prospect to detect ALPs that are produced in high-energy environments---in particular, core-collapse supernovae (CCSNe)---via the \mbox{Primakoff} resonant process \cite{Primakoff}, and subsequently travel undisturbed until they reach the Galactic magnetic field where they convert into $\gamma$-ray photons \cite{Fischer:2009af, Raffelt:1985nk}. For a 10-solar-mass (hereafter denoted by M$_{\odot}$) CCSN progenitor, the ALP spectrum should have a thermal shape peaking at around 70 MeV \cite{SN1987A_1, SN1987A_2, SN1987A_Payez}. The duration of an ALP-induced burst varies depending on the mass of a progenitor; nevertheless, the signal would be short (on the order of tens of seconds). No other physical processes are predicted to produce such spectral signatures in a CCSN's $\gamma$-ray spectrum. Thus, using observations of a CCSN and, in particular, searching for its presumed associated ALP-induced gamma-ray burst (GRB), can be an excellent probe for constraining the ALP parameter space (e.g. \cite{axionscope, Meyer:2020vzy}).
\par
Ordinary GRBs, believed to arise from collimated ultra-relativistic outflows of materials when, e.g., a star collapses, are among the most luminous events in the Universe, spectrally peaking in the keV--MeV energy range \cite{Piran_2005}. Depending on the duration of their prompt emission and their spectral hardness, GRBs are divided into two sub-types: the short-hard, for which the emission duration is less than 2 seconds; and long-soft, with their duration exceeding 2 seconds \cite{GRBs_1, GRB_2}. To explain differences between the two sub-types with respect to their duration, flux, variability, spectral parameters and evolution, the nature of their progenitors is often invoked \cite{GRB_Berger}. Short-hard GRBs are suspected to originate from compact-object binary mergers (such as two neutron stars or a neutron star and a black hole \cite{Fong_2013, Fong:2015oha, Monitor:2017mdv}) and long-soft GRBs are likely associated with Type Ib/c CCSNe \cite{1998Natur.395..670G, Patat:2001fn, Woosley:2006fn, 2007A&A...471..585B, 2011IJMPD..20.1745D, Cano:2016ccp}. Taking into account the predicted duration of an ALP-induced burst (a few tens of seconds), as well as the nature of the hypothesized ALP production site (CCSNe), we are particularly interested in studying the long-soft GRBs. 
\par
Using the properties of the ALP spectral emission, we first conduct a sensitivity analysis to determine the limiting distance to which \textit{Fermi} Gamma-ray Observatory \cite{GBM, LAT} would be able to detect an ALP-induced GRB using the LAT low-energy technique \cite{pelassa2010lat}. Considering astrophysical background levels from GRBs observed at various incidence angles, we estimate the necessary ALP flux that would lead to a significant detection of the ALP induced gamma-ray burst. Secondly, we consider a selected GRB sample and conduct a model comparison between fits that include the ALP spectral component and those that do not. Finally, we discuss the found limiting distances, the feasibility of upper limits on ALP couplings, and the tangibility of ALP detection with \textit{Fermi} or other gamma-ray observatories alike.
\par
This paper is organized as follows: \hyperref[sec:ALPmodel]{Section II} provides an overview of the ALP spectral model, derived in \cite{SN1987A_Payez}. In \hyperref[sec:dataselection]{Section III}, we describe the GRB data selection process. In \hyperref[sec:analysis]{Section IV}, we conduct a sensitivity study to determine the CCSN distances and photon-ALP couplings that would result in a significant detection of a GRB in the relevant MeV energy range with \textit{Fermi}. \hyperref[sec:GRBs]{Section V} describes the ALP-fitting method for the selected sample of GRBs. Finally, \hyperref[sec:discussion and future work]{Section VI} provides the summary and future outlooks for ALP searches within the gamma-ray energy band.

%% file: ALP_model.tex

\begin{figure*}
\centering
\includegraphics[width = 0.95\textwidth]{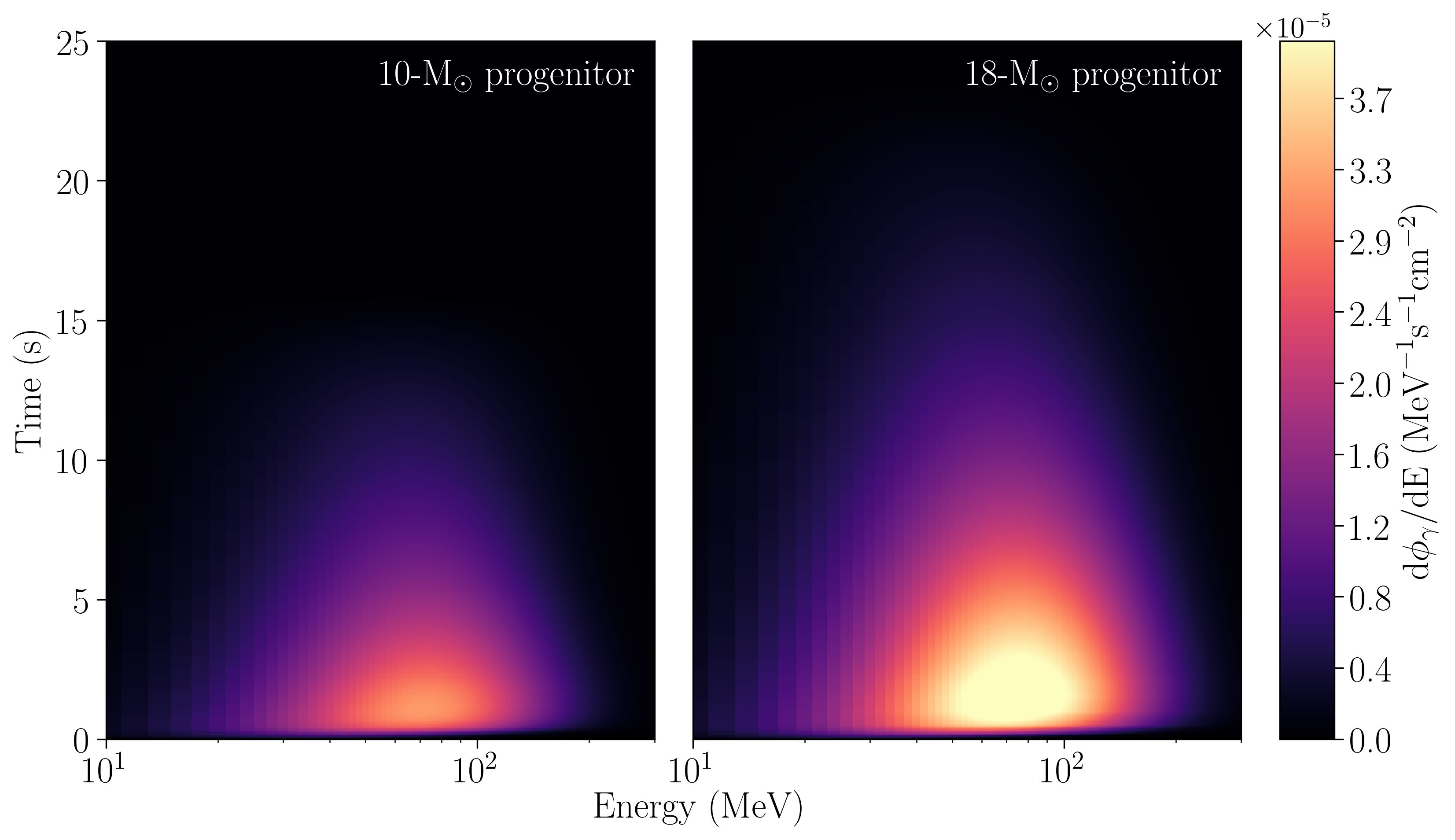}
\caption{Observed evolution of the  ALP-induced gamma-ray emission in time and energy in a core-collapse of a 10 and 18-M$_{\odot}$ progenitor, normalized by $N_{\textnormal{tot}} = 8.4 \times 10^{-54}$ cm$^{-2}$. Note that most of the emission occurs in the first ten seconds after the collapse. The 18-M$_{\odot}$ progenitor is a more energetic source of ALPs with a few-second prolonged emission as compared to the 10-M$_{\odot}$ progenitor.}
\label{fig:ALP_mesh}
\end{figure*}

To produce a spectral model for an ALP-induced GRB as observed on Earth, we utilize the one-dimensional CCSN (ALP-induced GRB) model derived in \cite{SN1987A_Payez} that is, due to the complexity of core-collapse modeling, available for only two distinct progenitor masses (10 and 18~M$_{\odot}$). The temporal and energy evolution of an ALP burst emission are shown in Fig.~\ref{fig:ALP_mesh}.

\begin{figure}[H]
\centering
\includegraphics[width = 0.48\textwidth]{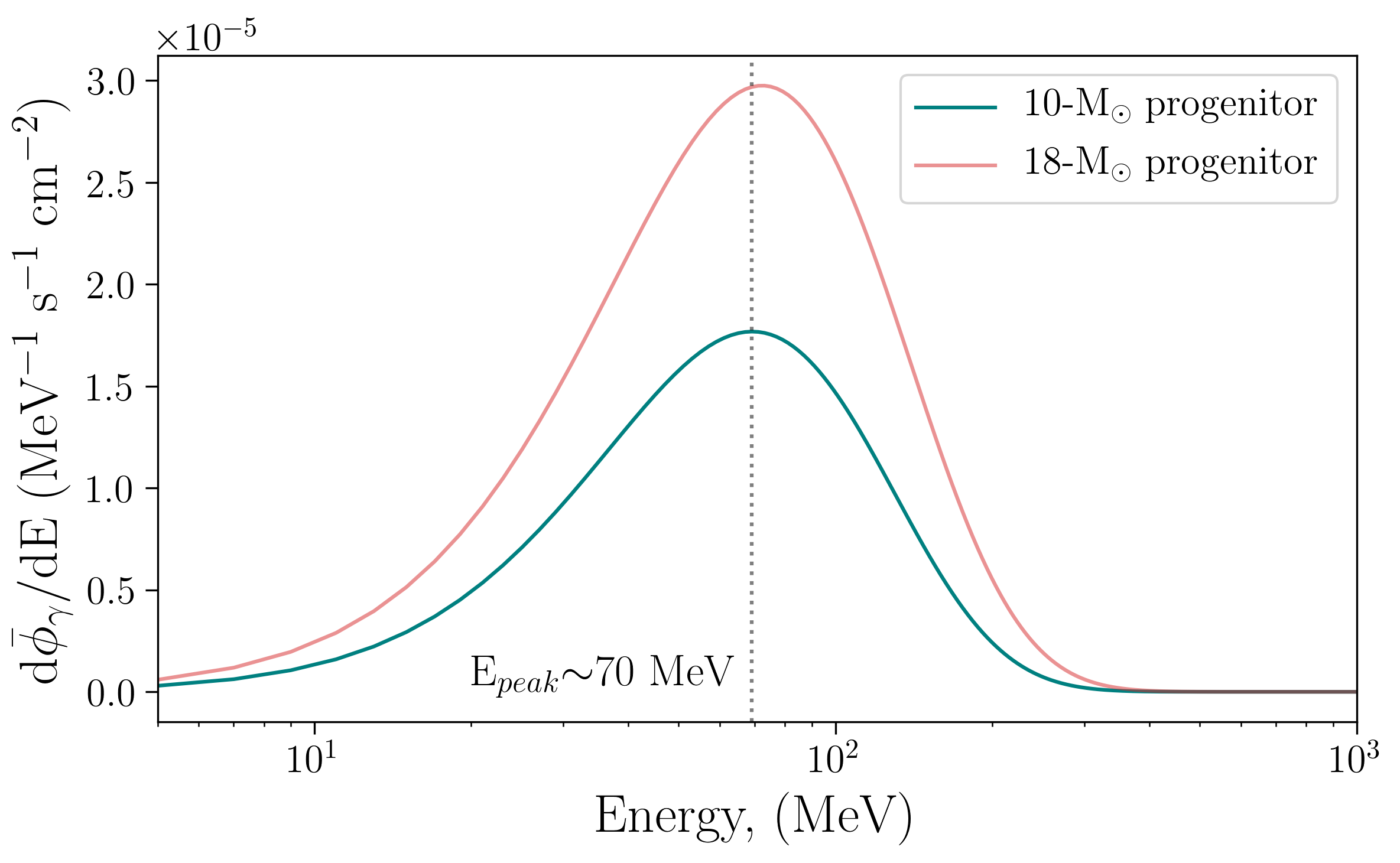}
\caption{The observed ALP-induced gamma-ray spectrum for 10 and 18-M$_{\odot}$ progenitors integrated and averaged over the first 10 seconds after the collapse for a normalization, $N_{\textnormal{tot}} = 8.4 \times 10^{-54}$ cm$^{-2}$ (see \hyperref[eqn:norm2]{Eq. 3}). Note that most of the flux is emitted around 70 MeV.}
\label{fig:ALPspec}
\end{figure}

The observed photon flux, $\dif  \phi_{\gamma} /\dif E$, can be expressed as
\begin{equation}
\label{eqn:obs_flux}
    \frac{\dif \phi_\gamma}{\dif E} = \frac{P_{a\gamma}(g_{a\gamma})}{4\pi d^2} \frac{\dif \dot{N_a}(g_{a\gamma})}{\dif E},
\end{equation}
where $d$ is the luminosity distance to the CCSN; $P_{a\gamma}$ is the ALP-photon conversion probability, proportional to $g_{a\gamma}^2$; and $ \dif \dot{N_a}/\dif E $ is the Primakoff production rate of ALPs per unit energy, also  proportional to $g_{a\gamma}^2$. This proportionality, $P_{a\gamma} \propto g_{a\gamma}^2$, breaks down before $P_{a\gamma}$ approaches unity \cite{Dobrynina:2014qba}. The total flux normalization may then be written as 
\begin{equation}
\label{eqn:norm2}
\begin{split}
    N_{\textnormal{tot}} = 8.4 \times 10^{-54} \textnormal{ cm}^{-2} \left(\frac{d}{10 \textnormal{ Mpc}}\right)^{-2} \times \left( \frac{g_{a\gamma}}{g_0}\right)^4 \\
    \times \left(\frac{P_{a\gamma}(g_0)}{0.1}\right),
\end{split}
\end{equation}
with $g_0 = 10^{-11}$ GeV$^{-1}$ denoting an arbitrary reference coupling, roughly corresponding to the current upper limit, $g_{a\gamma} \lesssim 5.3\times 10^{-12}$~GeV$^{-1}$ \cite{SN1987A_Payez}, for ALP masses ranging from $m_a\simeq10^{-12}$ to $10^{-8}$~eV
\cite{SN1987A_Payez, TheFermi-LAT:2016zue, Malyshev:2018rsh, Cheng:2020bhr, Buehler:2020qsn}. For lower-mass ALPs, i.e.\ $m_a \lesssim 10^{-12}$~eV, observations of galaxy clusters provide more stringent constraints, $g_{a\gamma} \lesssim 6-8\times10^{-13}$~GeV$^{-1}$ \cite{Reynolds:2019uqt} (see also \cite{Libanov:2019fzq} and \cite{Dessert:2020lil, Buen-Abad:2020zbd}). Furthermore, for masses $m_a \gtrsim 10^{-10}$~eV, the conversion probability $P_{a\gamma}$ becomes energy-dependent and effectively drops within the MeV energy range considered in this analysis. The observed photon flux can now conveniently be expressed as
\begin{equation}
\label{eqn:obs_flux}
   \frac{\dif \phi_\gamma}{\dif E} = N_{\textnormal{tot}} \times \frac{\dif \dot{N_{a}}(g_0) }{\dif E}.
\end{equation}
Using the CCSN model in \cite{SN1987A_Payez}, we obtain the temporal and energy information about the ALP production rates ${\dif \dot{N_{a}}(g_0) }/{\dif E}$ in a core-collapse due to the ALP interactions described in Eq.~\eqref{eq:L-alp}. Figure~\ref{fig:ALP_mesh} shows that most of the corresponding ALP-induced gamma-ray emission happens in the first few tens of seconds for both progenitor masses; hence, by averaging over the time interval of 10 seconds starting at the core-collapse, we obtain the expected spectra shown in Fig.~\ref{fig:ALPspec}. 

\subsection{Conversion Probability}
\label{subsec:conversionP}
\begin{figure}[]
\centering
\includegraphics[width = 0.5\textwidth]{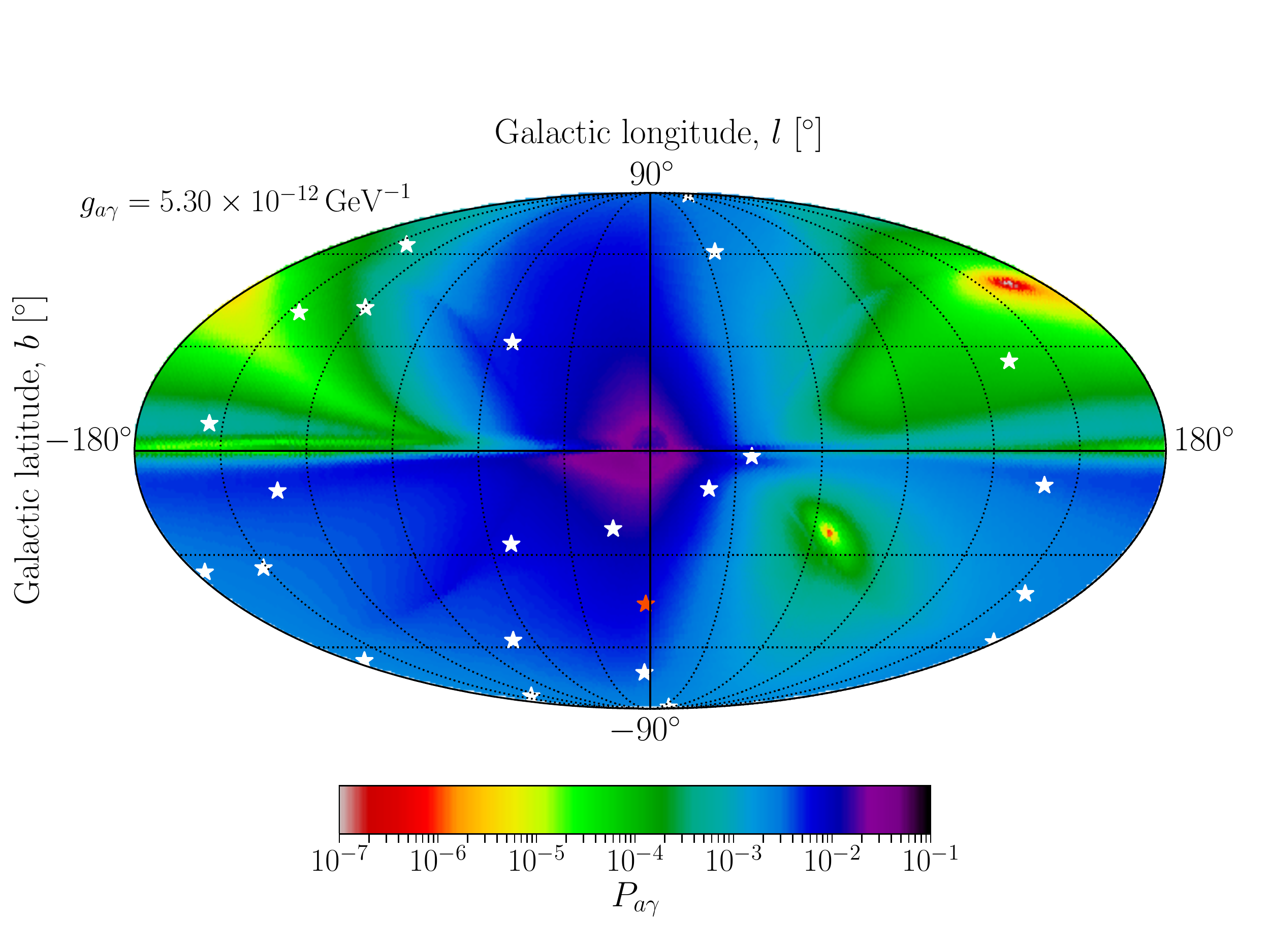}
\caption{ALP-photon conversion probability as a function of source position in the sky. We consider only the ALP conversion into gamma rays within the Milky Way's coherent magnetic field component, as modeled in \cite{axionscope}. We assume an ALP mass m$_a = 10^{-10}$ eV and coupling g$_{a\gamma}=5.3 \times 10^ {-12}$ GeV$^{-1}$, and energy $10<E<300$~MeV. Note that for this configuration, $P_{a\gamma, \text{max}} \sim 0.1$ is the maximum conversion probability reached in the central regions of Milky Way. White crosses represent the best localization positions of the GRB sample considered in Sec.~\ref{sec:GRBs}, with the red cross corresponding to GRB 101123A \cite{Ajello_2019}.}
\label{fig:conversion}
\end{figure}

The photon-ALP conversion probability, $P_{a\gamma}$, is computed numerically to account for variations in the Galactic magnetic field. Following the Milky Way magnetic field model by  Jansson $\&$ Farrar \cite{J&Ferrar}, we compute the conversion probabilities for different positions in the sky, assuming that the photon-ALP mixing happens throughout the entire Galaxy, as done in \cite{axionscope, Manuel_code}. The contribution from the turbulent magnetic field component is not included in this analysis since its typical coherence length ($\sim$10 pc) is significantly shorter than the ALP-photon oscillation length and can be neglected for the considered ALP mass $m_a \lesssim 10^{-10}$~eV in most sky regions \cite{Meyer_2013, Carenza:2021alz}. For an emission that passes only through the Galactic magnetic field, the conversion probability is shown in Fig.~\ref{fig:conversion}, as a function of source's position in the sky. We assume that $P_{a\gamma}$ is energy-independent in our analysis, which is valid for low-mass ALPs
($m_a \lesssim 10^{-10}$~eV), while for larger ALP masses $P_{a\gamma} = P_{a\gamma}(g_{a\gamma},m_a,E)$ decreases and starts to oscillate as a function of energy \cite{SN1987A_Payez}. 
\par 
We do not take into account the photon-ALP conversions that may happen in the intergalactic magnetic field, as such contributions would be negligible and highly uncertain for the nearby sources we consider. For example, if we assume a uniform intergalactic magnetic field strength of $\sim$1 nG, a coupling of $g_{a \gamma} = 5.3\times 10^{-12}$ GeV$^{-1}$ and a distance of 5~Mpc, we obtain that in the strong-mixing regime (here reached for $m_a \lesssim 10^{-11}$~eV) \cite{Mirizzi:2009aj}, the order of the conversion probability, $\mathcal{O}(P_{a\gamma})$, is $10^{-3}$. Furthermore, considering the conversion probability in an extragalactic source's host galaxy and, if applicable, its surrounding intracluster medium, would be highly uncertain due to a lack of knowledge of their respective magnetic fields and would have to be taken on a case by case basis. In addition, such consideration would result in an increase of the observed gamma-ray flux \cite{axionscope}, resulting in adjustments to Fig.~\ref{fig:conversion} with  conversion probabilities unlikely reaching values below $10^{-3}$; hence, neglecting them renders our results conservative. Due to case by case differences,  we do not take into consideration the ALP-photon conversion that may take place in the magnetic field of the intergalactic medium and within the host galaxy.

\subsection{ALP Model in \textit{XSPEC}}
\label{subsubsec:ALP-xspec}

Spectral modeling in this paper is conducted using the standard high-energy fitting package \textit{XSPEC} \cite{XSPEC_code} and its Python adaptation, \textit{PyXspec} \footnote{\url{https://heasarc.gsfc.nasa.gov/docs/xanadu/xspec/python/html/index.html}, accessed on April 23, 2019.}. The ALP spectra shown in Fig.~\ref{fig:ALPspec} are used to write a model function that is inserted into the \textit{XSPEC} model library using the \texttt{addPyMod} method. For the model parameters, we consider two different progenitor masses, 10 M$_{\odot}$ and 18 M$_{\odot}$, with the normalization parameter $N_{\textnormal{tot}}$, from Eq.~\eqref{eqn:obs_flux}, left free to vary.

%% file: data_selection.tex

\subsection{\Fermi observatory}
\label{subsec:Fermi}

The \textit{Fermi} observatory provides a wide spectral coverage and excellent sensitivity for studying GRBs. The observatory contains two instruments on board, the Gamma-ray Burst Monitor (GBM, \cite{GBM}) and the Large Area Telescope (LAT, \cite{LAT}). The GBM has twelve sodium-iodide (NaI) and two bismuth-germanate (BGO) scintillation detectors, covering 8~keV to 1~MeV and 150~keV to 40~MeV in energy, respectively. The field of view (FoV) is $\sim$9.5~sr and the point-source localization accuracy is $\sim$5$\degree$. On the other hand, LAT is a pair production telescope covering the energy range from 20~MeV to  more than 300~GeV, with a FoV of $\sim$2.4~sr, and a point-source localization of $<$1$\degree$, for energies above 1~GeV. Figure \ref{fig:conversion} shows the best localization values for the considered GRB sample using either GBM or LAT instruments \cite{Ajello_2019}. Particularly of interest in this paper is the LAT low-energy data (LLE, \cite{pelassa2010lat}), due to the energy of the ALP spectral peak at 70~MeV, shown in Fig.~\ref{fig:ALPspec}.
\par
The LLE analysis method was developed with the goal of maximizing the effective area of the LAT instrument in the low-energy regime \cite{pelassa2010lat}. This is done by relaxing the requirements on background rejection, as compared to the standard LAT transient analysis. This technique is particularly useful for a study of transients, such as  GRBs, for energies greater than $\sim$30~MeV. The LLE event selection relies upon having at least one reconstructed track within the LAT's tracker/converter (TKR), allowing for an estimate of the direction of the incoming photon. Furthermore, all photons pass through the anticoincidence detector (ACD) which enables cosmic-ray background rejection. Finally, this algorithm requires a non-zero reconstructed energy of the considered event. Then, for the short and bright transient sources, the background is determined by an ``ON" and ``OFF" time-interval technique. The background rate during the ``OFF" interval is fit by a polynomial function in each energy bin, providing us with an estimate of the background during the ``ON" interval. The corresponding LLE response files are produced using Monte Carlo simulations of bright point sources with a specific spectral shape at the sky position of interest. The systematic effects in reconstructing the LLE events are considered in \cite{Fermi-LAT:2013cla}, estimating the discrepancy between the LLE selection criteria in the LAT data and in Monte Carlo simulations to be $\sim17\%$ for events below 100~MeV. Comparing the flux values reveals that LLE's flux estimations are on average lower than those from the standard LAT analysis; however, no significant biases are reported for the energy resolution, with that of LLE estimated to $\sim$40\% at 30~MeV and $\sim$30\% at 100~MeV. A detailed description of the LLE technique is provided in \cite{Fermi-LAT:2013cla}.
The LLE data is publicly available from the HEASARC website \footnote{\url{https://heasarc.gsfc.nasa.gov/W3Browse/fermi/fermille.html}, as accessed on December 24, 2019.}. In Sec. \ref{sec:analysis}, we focus on utilizing the LLE data sample; which, in Sec. \ref{sec:GRBs}, is complemented by the GBM and the standard LAT transient data, when such observations are available.

\subsection{Time-tagging of a core collapse}
\label{subsec:time-tag}

One of the main challenges for conducting ALP spectral fitting is to observationally approximate the collapse time of a supernova, i.e.\ the time when most of the ALPs escape the collapse site. 
\par
The optimal way to address this challenge is 
through neutrino detection from the source, as neutrinos and ALPs are expected to arrive at approximately the same time. However, observational neutrino data from supernovae are scarce (so far, only SN1987A \cite{SN1987A_Payez}). Although the second generation of neutrino detectors has significantly improved in sensitivity (e.g.\ IceCube detection from the blazar TXS 0506+056 at $z = 0.33$ \cite{IceCube}), no neutrino signal detection is expected from an extragalactic supernova in the near future \cite{Ando:2005ka}. This imposes tight limitations on a potential ALP-source distance from which a neutrino signal can be detected, likely to a CCSN in our own Galaxy or in the Local Group, extending up to a few Mpc \cite{Ando:2005ka,Kistler_2011}. Furthermore, if we were to use neutrinos for time-tagging of a core collapse, we would also require a gamma-ray observation of such an event in order to conduct the ALP spectral fitting. For example, the probability for a Galactic SN to occur in the LAT FoV in the next 3 years is $\lesssim$1~\%.
\par
Beside neutrino detection, another way to approximate explosion times of supernovae is by using their optical lightcurves \cite{optical}. This technique has been used in \cite{Meyer:2020vzy} to search for an ALP-induced gamma-ray burst with the standard LAT data above 60 MeV. 
\par
Another possibility to infer the core collapse time is from the time of the ordinary astrophysical GRB. The ordinary bursts are delayed on the order of seconds to minutes with respect to the core collapse, as the jet needs to form and propagate through the stellar envelope (see, e.g., Fig.~10.1 in \cite{woosley2011models}). Moreover, the ALP-induced gamma-ray emission is approximately isotropic from the source, in contrast to the ordinary GRB jet emission---which also might not be aligned with our line of sight, resulting in a considerably weaker signal if seen off-axis (or even a `failed GRB' \cite{Huang:2001hj}). This could imply that not every ALP signal is accompanied by a subsequent ordinary GRB detection. A dedicated study regarding precursor emission (hypothetically, an ALP signal prior to the observed jet emission) to GRBs may address the time-tagging issue in more detail and is a matter for future research. In this paper, we assume that the time when most ALPs escape the collapse site coincides with the GRB signal time window. 

\subsection{GRB Selection Criteria}
\label{subsec:selection}

Considering the GRBs detected so far by \textit{Fermi}-LAT \cite{Ajello_2019}, with their corresponding optical follow-ups and, in turn, redshift information, we infer that all associated sources are too far (the closest one being over 600~Mpc away) to be considered for a sizable ALP-induced burst observation (see the sensitivity results of Sec.~\ref{subsec:sens}).
Thus, in our analysis we instead only consider all the LLE-detected GRBs without redshift information (hereafter referred to as \textit{unassociated} GRBs) as potential ALP signal candidates; albeit, most likely ordinary GRBs of extragalactic origin. With limited information on the origin of the considered GRB sample, we assume they are either induced by an ALP signal or by ordinary astrophysical processes traditionally applied in GRB spectral modeling (see, e.g., \cite{Ajello_2019}). This allows us to start the ALP analysis at the GRB trigger time, T$_0$, with the considered time window encompassing either (or both) the traditional GRB emission and the potential ALP signal.
\par
We consider all unassociated GRB detections by \textit{Fermi} LAT from August 2008 to August 2018, as reported in the \textit{Fermi} LAT Second Gamma-Ray Burst Catalog, 2FLGC \cite{Ajello_2019}, publicly available on the HEASARC website \footnote{\url{https://heasarc.gsfc.nasa.gov/W3Browse/fermi/fermilgrb.html}}.  Motivated by the energy of the ALP spectral peak at $\sim$70 MeV, we further restrict our sample to GRBs with at least 5$\sigma$ detection in LLE alone. Such strong signal in the low-sensitivity region of LAT often indicates a strong signal in either GBM, or standard LAT (or, both), often meriting follow-up observations by optical telescopes. Thus, the 5-$\sigma$ requirement combined with no redshift information are the most exclusive cut criteria for our sample. We require the source to be within the FoV throughout the entire duration of the burst which, required by the nature of ALP emission, should be a long GRB. For the sensitivity analysis (Sec.~\ref{sec:analysis}), in order to 
somewhat increase our 
GRB background sample, we drop the GRB duration criterion as we are only concerned with the observed background levels, which are independent from the GRB's duration. This results in three additional short GRBs (GRB 081024B, GRB 090227B, and GRB 110529A) which can be used as background templates.
As such, the spatial distribution of all the considered GRBs, as shown in Fig.~\ref{fig:conversion}, forms a representative sample of the GRB sources in the sky. 
\par 
The GRB selection criteria are summarized in Table~\ref{tab:selection}. From the initial sample of 186 LAT-detected GRBs, applying the above selection cuts results in a sample of 24 long GRBs, all listed in Table~\ref{tab:GRBlist}. 

\begin{table}[]
\centering
\caption{Selection criteria applied to GRBs detected by \textit{Fermi}-LAT between August 2008 to August 2018. From the initial sample size of 186 GRBs, we narrow down the list of candidates to 24 shown in Table~\ref{tab:GRBlist}. The burst duration criterion is not applied to the sensitivity analysis in Sec.~\ref{sec:analysis}, allowing for the inclusion of three additional short GRBs.}
\begin{tabular}{ll}
\hline \hline
Property               & Selection Criterion \\
\hline
Distance               &  unassociated (no redshift)                  \\
Detection significance &     $\geq 5\sigma$ in LAT-LLE ($\gtrsim 30$ MeV)               \\
Observed time interval          &  $\geq$ duration of the GRB\footnote{We select only the GRBs that are within LAT's FoV throughout their entire duration. Furthermore, we consider a few-hundred-second padding before and after the burst duration for modeling the background emission.}   \\
Burst duration & long GRBs ($T_{95}\gtrsim 2$ seconds\footnote{as reported in Table III in \cite{Ajello_2019}}) \\
    & \textit{(not used in Sec.~\ref{sec:analysis})}\\
\hline
\hline
\end{tabular}
\label{tab:selection}
\end{table}

\begin{table}[]
\caption{List of the 24 GRBs that pass the selection criteria. $T_{95}$ corresponds to the duration reported in Table III in \cite{Ajello_2019}, as seen by GBM. The following columns show the best-fit models listed without an additional ALP component, with uncertainties representing the 90-\% confidence interval for the given fit parameter. Also included are the log-likelihood ratio (LLR) $\Lambda$ values,  $\Lambda= -2 \log \left(L_{\textnormal{GRB}}/L_{\textnormal{GRB+ALP}}\right)$, which are derived in Sec.~\ref{sec:GRBs}. We also report the best-fit parameters for the cases in which the Band model (denoted by \textit{XSPEC}'s \texttt{grbm}) is one of the model components \cite{1993ApJ...413..281B}, to demonstrate that the parameters do not reproduce the ALP spectral shape which may be reproduced with $\alpha_1 \sim -2.4, \alpha_2 \sim -0.1,$ and $E_c \sim 30$ MeV. Details of the GRB analysis are described in Sec.~\ref{sec:GRBs}.}
\resizebox{0.48\textwidth}{!}{
\begin{tabular}{lllcccr}
\hline \hline

GRB  & T$_{95}$  & Best model & \multicolumn{3}{c} {\texttt{grbm} parameters} & LLR\\
 & (s) & (no ALP) & $\alpha_1$ & $\alpha_2$ & $E_\text{c}$ (keV)& \\
\hline

080825C & 22.2 & \texttt{grbm} & -0.65 $\substack{+0.05 \\ -0.05}$ & -2.41 $\substack{+0.04 \\ -0.04}$ & 143$\substack{ +13\\-12 }$  & 0.2 \\

090217  & 34.1 & \texttt{grbm} & -1.11 $\substack{+0.04\\ -0.04}$ & -2.43 $\substack{+0.03 \\ -0.04}$ & 16$\substack{ +13\\-8 }$  & 0.1\\

100225A & 12.7 & \texttt{grbm} &-0.50$\substack{+0.25 \\ -0.21}$ & -2.28$\substack{+0.07 \\ -0.09}$ & 223$\substack{+112 \\ -68}$ & 0.0 \\

100826A & 93.7  & \texttt{grbm + bb} & -1.02$\substack{+0.04 \\ -0.04}$ & -2.30$\substack{+0.03 \\ -0.04}$ & 484$\substack{+72 \\ -63}$  & 0.0\\

101123A & 145.4  & \texttt{grbm + cutoffpl} & -1.00$\substack{+0.07 \\ -0.08}$ & -1.94$\substack{+0.15 \\ -0.12}$ & 187$\substack{+74 \\ -62}$ &    5.8\\

110721A & 21.8 & \texttt{grbm + bb} & -1.24$\substack{+0.02 \\ -0.01}$ & -2.29$\substack{+0.03\\-0.03}$ & 1000$\substack{+28 \\ -39}$ & 0.0\\

120328B & 33.5 & \texttt{grbm + cutoffpl} & -0.67$\substack{+0.06 \\ -0.05}$ & -2.26$\substack{+0.05 \\ -0.05}$ & 101$\substack{+12\\-13}$ & 0.0\\

120911B & 69.0     & \texttt{grbm} & -2.50$\substack{+0.92 \\ -1.04}$ & -1.05$\substack{+0.63 \\ -0.38}$ & 11$\substack{+10 \\ -2 }$ & 0.0\\

121011A & 66.8 & \texttt{grbm} & -1.08$\substack{+0.10 \\ -0.21}$ & -2.18$\substack{+0.11 \\ -0.16}$  & 997$\substack{+84 \\ -26}$  & 0.0 \\

121225B & 68.0 & \texttt{grbm} &  -2.38$\substack{+1.02 \\ -0.40}$ &  -2.45$\substack{+0.06 \\ -0.07}$ & 11$\substack{+89 \\ -3}$ & 0.0   \\

130305A & 26.9 & \texttt{grbm} &  -0.76$\substack{+0.03 \\ -0.03}$ &  -2.63$\substack{+0.06\\ -0.06}$ & 665$\substack{+61 \\ -55}$  & 0.0 \\

131014A & 4.2   & \texttt{grbm} &-0.55$\substack{+0.33 \\ -0.98}$ & -2.65$\substack{+0.17 \\ -0.19}$ &  255$\substack{+36 \\ -11}$ & 0.63 \\

131216A & 19.3   & \texttt{grbm + cutoffpl} & -0.46$\substack{+0.28\\-0.24}$ & -2.67$\substack{+1.94 \\ -0.94}$ &  178$\substack{+77 \\ -92}$ & 0.0 \\

140102A & 4.1   & \texttt{grbm + bb} & -1.10$\substack{+0.12\\-0.09}$ & -2.41$\substack{+0.16 \\ -0.11}$ &  206$\substack{+65 \\ -92}$ & 2.3 \\

140110A & 9.2   & \texttt{grbm} & -2.49$\substack{+1.64 \\ -1.59}$ & -2.19$\substack{+0.20 \\ -0.22}$ &  11$\substack{+23 \\ -3}$ &  0.0 \\

141207A & 22.3 & \texttt{grbm + bb} & -1.21$\substack{+0.09 \\ -0.06}$ &  -2.33$\substack{+0.11 \\ -0.13}$ &  999$\substack{+18 \\ -70}$ &    0.0  \\

141222A & 2.8  & \texttt{grbm + pow} & -1.57$\substack{+0.03 \\ -0.02}$ & -2.83$\substack{+0.46 \\ -1.74}$ & 9971$\substack{+390 \\ -832}$ &  0.0     \\

150210A & 31.3  &  \texttt{grbm + pow} & -0.52$\substack{+0.04 \\ -0.05}$ & -2.91$\substack{+0.11 \\ -0.38}$ & 1000$\substack{+517 \\ -234}$ &  0.0    \\

150416A & 33.8 &  \texttt{grbm} & -1.18$\substack{+0.04 \\ -0.04}$ & -2.36$\substack{+0.13 \\ -0.21}$ & 999$\substack{+187 \\ -269}$&  0.0 \\

150820A & 5.1  & \texttt{grbm} & -0.99$\substack{+0.56 \\ -1.30}$ & -2.01$\substack{+0.82 \\ -0.27}$ & 303$\substack{+61 \\ -39}$ & 0.0  \\

151006A & 95.0  &  \texttt{grbm} & -1.35$\substack{+0.06 \\ -0.03}$ &  -2.24$\substack{+0.07 \\ -0.08}$ & 998$\substack{+33 \\ -84}$   & 0.0\\

160709A & 5.4   & \texttt{grbm + cutoffpl} &  -1.44$\substack{+0.18\\ -0.12}$ &  -2.18$\substack{+0.15 \\ -0.18}$ &  9940$\substack{ +373 \\ -511}$  & 1.0 \\

160917A & 19.2   & \texttt{grbm + bb} &  -0.78$\substack{+3.45 \\ -1.40}$ &  -2.39$\substack{+0.20 \\ -0.10}$  & 994$\substack{+634 \\ -216}$ & 0.9 \\ 

170115B & 44.8    & \texttt{grbm} &  -0.80$\substack{+0.02 \\ -0.04}$ &  -3.00$\substack{+0.10 \\ -0.07}$ & 1000 $\substack{+226 \\ -106}$ & 2.8 \\ 
\hline
\hline
\end{tabular}}
\label{tab:GRBlist}
\end{table}

%% file: sensitivity.tex
We conduct a sensitivity study using the LLE data for two reasons: firstly, to ensure that a manually injected ALP feature can be recognized by our fitting algorithm; secondly, to determine the maximum distance for a given photon-ALP coupling for which the ALP feature is still significantly detectable in LLE. 
We also find the distance to a CCSN that allows for setting competitive upper limits on the photon-ALP coupling with \textit{Fermi} LLE. In this study, we use a background energy spectrum derived from three different GRB observations and manually inject the ALP-induced gamma-ray signal that is subsequently folded with the instrument response function. 

\subsection{Background Considerations}
\label{subsec:bkg}
\par
We extract background information for each individual GRB using the analysis tool developed by the LAT team, \textit{gtburst}. This tool allows for a selection of the off-signal intervals, which are fitted in each energy channel with a polynomial function in time, resulting in a fitted background count rate. A detailed description of the GRB analysis process with \textit{gtburst} may be found in Sec.~\ref{sec:GRBs}. 
\par
\begin{figure}[]
\centering
\includegraphics[width = 0.5\textwidth]{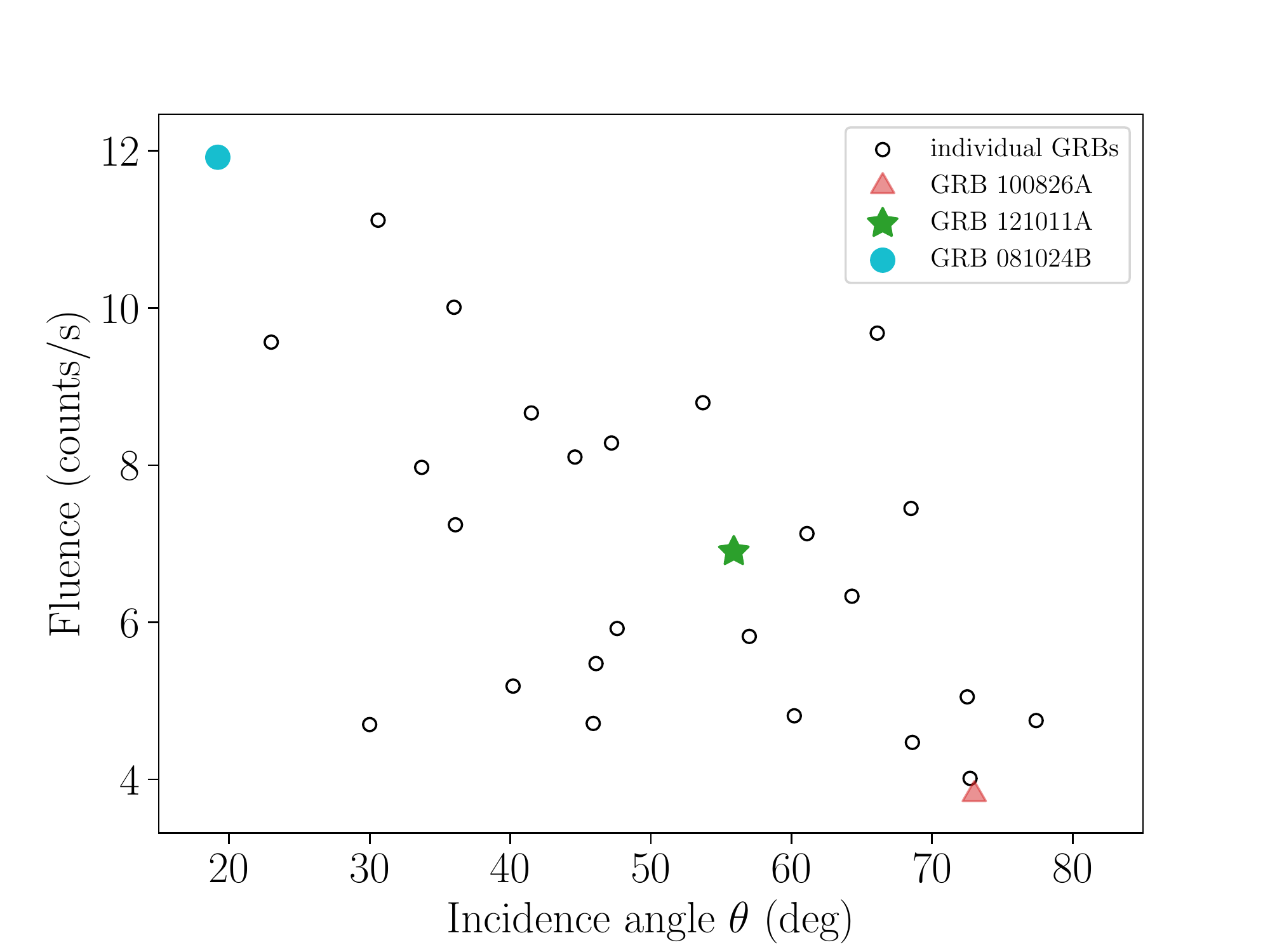}
\caption{Background fluences plotted against the incidence angles to the detector, $\theta$. Pink triangle, green star, and blue circle respectively correspond to the lowest, median, and highest background fluences in the sample. Note, however, that low count numbers may be caused by a significant drop in LLE effective area at high $\theta$s.}
\label{fig:bkg}
\end{figure}

Once we obtain the background count rates, we compute their fluences by conducting spectral fitting using a power-law model, half-Gaussian profile, or a combination thereof to find the lowest, highest, and typical values of background fluences for the LLE data selection, for energies starting at $\sim 30$ MeV and reaching the GeV energies. In addition to the GRB sample listed in Table~\ref{tab:GRBlist}, we include three supplementary short GRBs that pass the remaining selection criteria, GRB 081024B, GRB 090227B, and GRB 110529A. The found fluences for GRB backgrounds are computed and plotted as a function of the incidence angle $\theta$, shown in Fig.~\ref{fig:bkg}. 
It is important to note that for large incidence angles, LLE effective area drops significantly, resulting in low count rates. In our sensitivity study, we therefore restrict our analysis to GRBs that are less than $\sim$70 degrees off-axis, and conduct a study of the median, the lowest and the highest background fluence values. For these GRBs, the lower background count rates reflect a drop in the effective area for LLE events, and \textit{should therefore not be interpreted as GRBs with necessarily lower gamma-ray backgrounds.} Finally, as long GRBs are expected to be uniformly distributed across the sky \cite{Ajello_2019}, we may assume a representative ALP-photon conversion probability $P_{a\gamma}$, shown in Fig.~\ref{fig:conversion}, for each considered background in our analysis (albeit, the GRB distribution may be  anisotropic if the unassociated sources are very close-by.)

\subsection{Simulating the ALP spectrum}
\label{subsubsec:ALPsimulation}
\begin{figure}
\centering
\includegraphics[width = 0.5\textwidth]{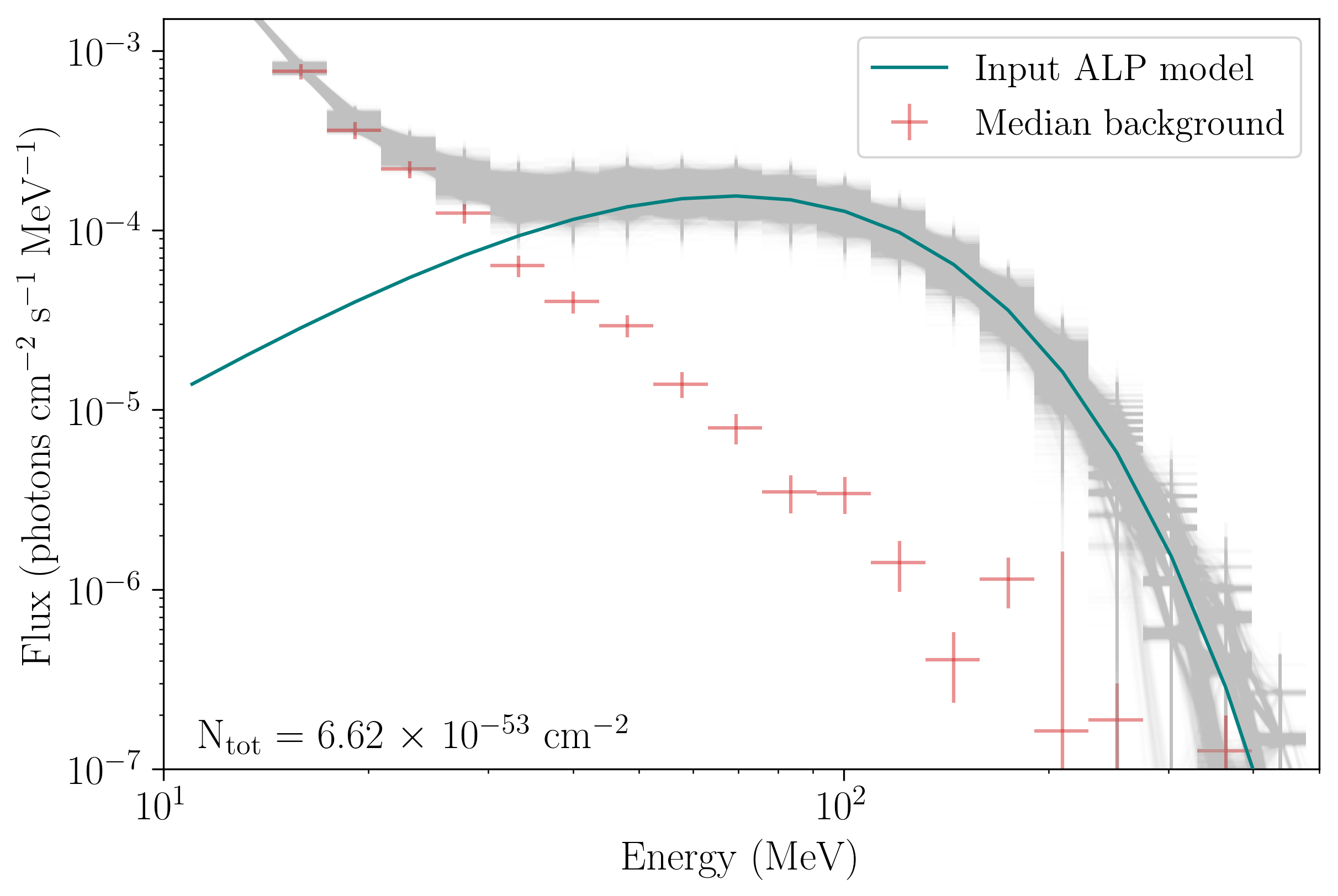}
\caption{\textit{XSPEC} \texttt{fakeit} ALP simulations \cite{XSPEC_code}. Shown in gray is a sample of 2000 realizations of the ALP spectrum for a 10-M$_{\odot}$ progenitor and normalization $N_{\text{tot}}= 6.62 \times 10^{-53}$ cm$^{-2}$ (corresponding to a $\sim$1-Mpc distant CCSN with $P_{a\gamma}=0.1$ and $g_{a\gamma}=5.3\times10^{-12}$~GeV$^{-1}$ \cite{SN1987A_Payez}) on top of the median background, including the background statistical and systematic uncertainties produced with \textit{gtburst}. The solid blue line represents the input ALP signal for the quoted normalization value.}
\label{fig:simulation}
\end{figure}

To simulate ALP-induced gamma-ray spectra, we use the \textit{XSPEC}'s \texttt{fakeit} function. We consider the ALP spectra in the energy range given by the LLE data file specifications, starting at $\sim$30~MeV and reaching the GeV energies. We use the response function and the background observation derived previously from each considered LLE-detected GRB. On top of the scaled ALP-induced gamma-ray signal from Fig.~\ref{fig:ALPspec}, we add a realization of the background, taken to be a power-law approximation to the channels' photon rates fits for each considered GRB. The combination of the signal and the background is then passed through the \textit{XSPEC}'s \texttt{fakeit} function to create 2000 realizations of spectra corresponding to different normalization values of the ALP signal on top of the observed background levels. An example of a simulation sample resulting from  the \textit{XSPEC}'s \texttt{fakeit} function is shown in Fig.~\ref{fig:simulation}.

\subsection{Sensitivity results}
\label{subsec:sens}

\begin{figure*}[]
\centering
\subfloat{\includegraphics[width = 0.5\textwidth]{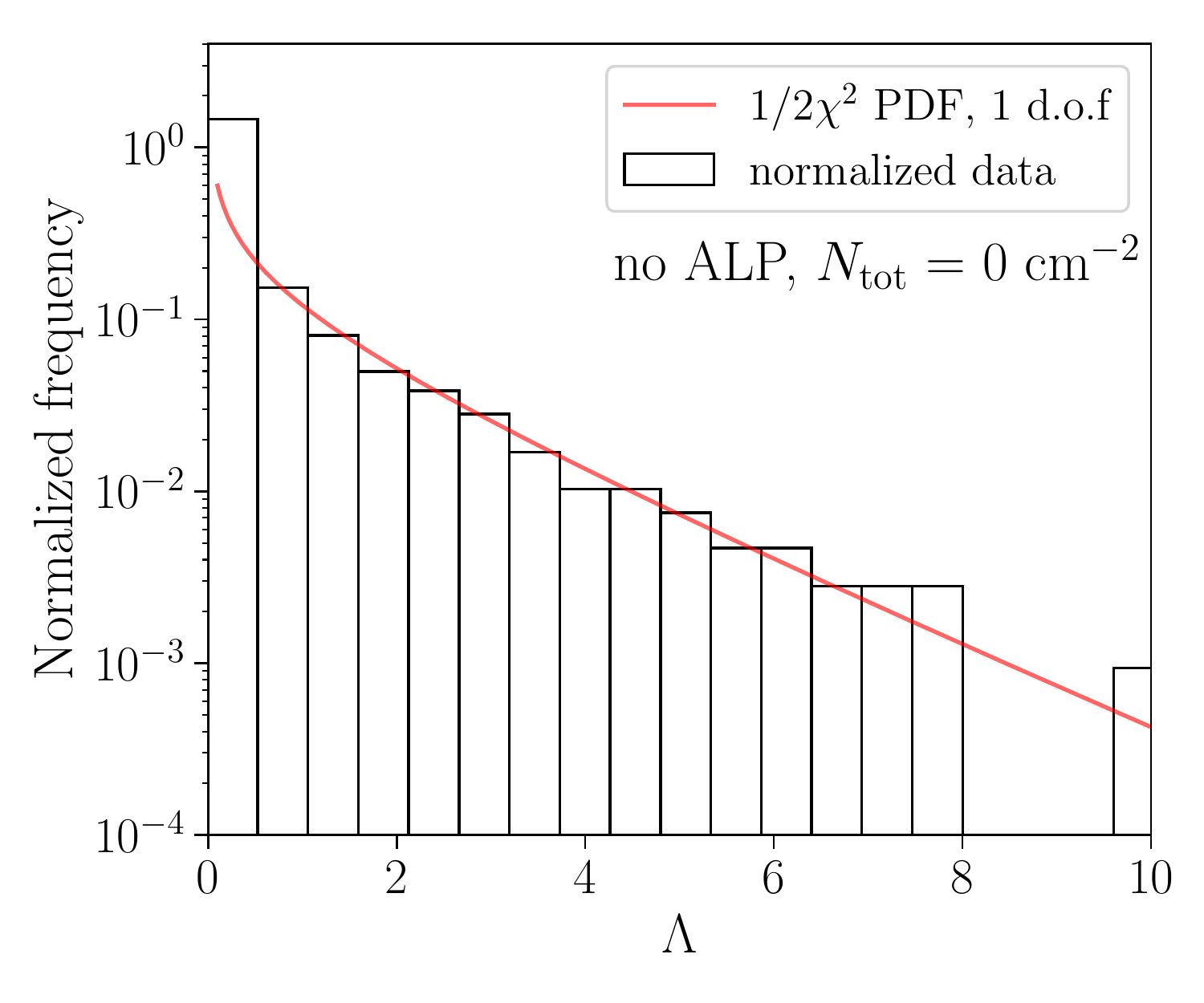}}
\subfloat{\includegraphics[width = 0.5\textwidth]{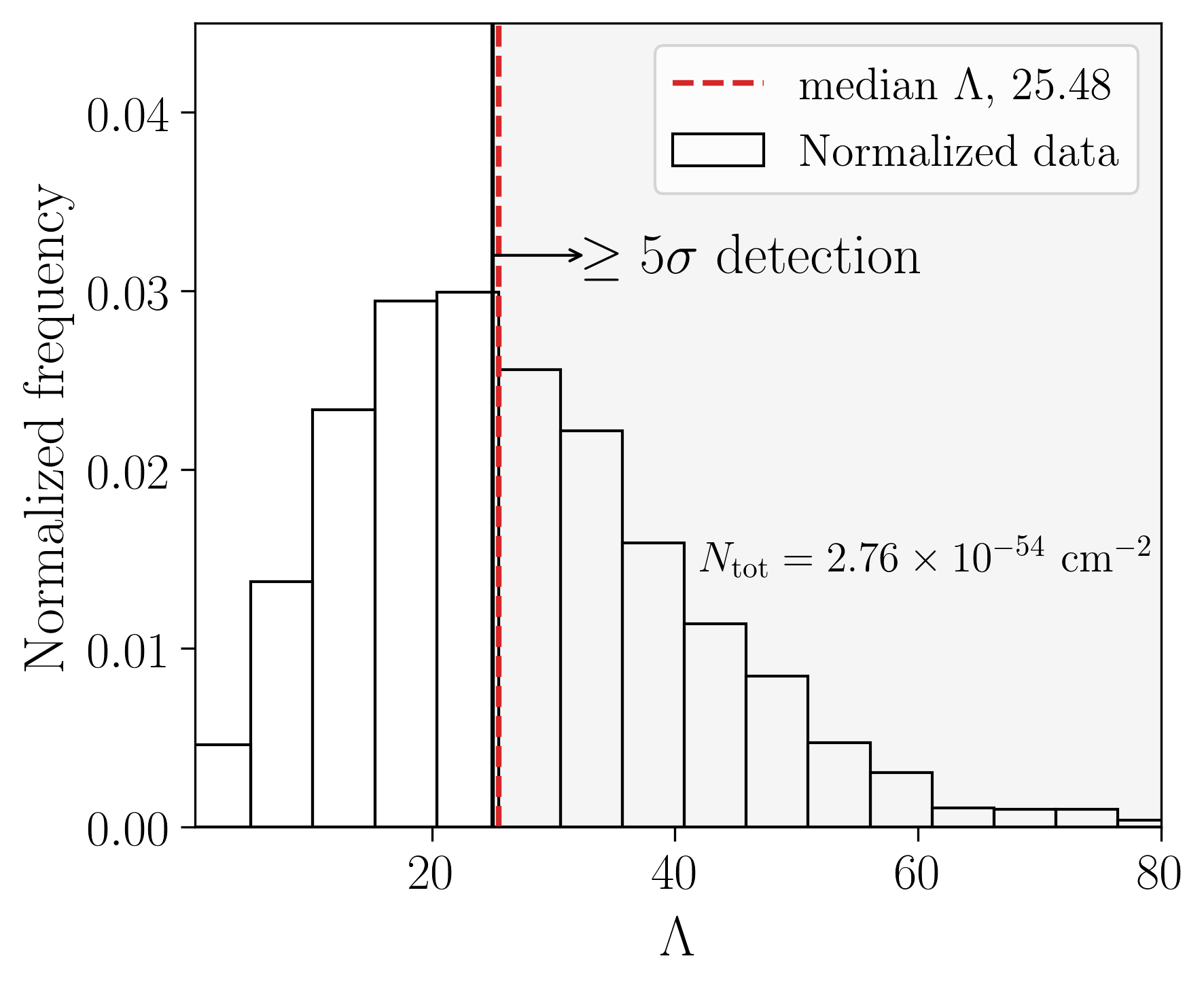}}
\caption{Left:  demonstration of Wilks' theorem for the considered simulated spectra \cite{wilks1938, Cowan:2010js}. The red line is the half-$\chi^2$ distribution with 1 d.o.f. on top of our normalized $\Lambda$ distribution when no additional ALP signal is injected to the median background. 
Right: the distribution of the same $\Lambda= -2 \log \left(L_{\textnormal{null}}/L_{\textnormal{alternative}}\right)$ when an additional ALP signal is injected to the median background. The null model remains the one without an additional ALP component, but the realizations are now drawn from a model with $N_{\textnormal{tot}} = 2.76 \times 10^{-54}$ cm$^{-2}$, for the case of the median background and a progenitor mass of 10 M$_\odot$. The median $\Lambda$ represents the median value of the histogram. We determine the sensitivity of our experiment to a deviation from a no-ALP scenario, corresponding to a threshold of 5$\sigma$ or the $p$-value of $5.7 \times 10^{-7}$.}
\label{fig:Wilks}
\end{figure*}

To find the \textit{Fermi}-LAT sensitivity to detecting ALP-induced gamma-ray signal originating from a given CCSN using the LLE data, we consider the highest, the lowest, and the median background levels as seen in our GRB sample, respectively corresponding to backgrounds of GRB 081024B, GRB 100826A, and GRB 121011A (also corresponding to the low, high, and medium $\theta$s; see Fig.~\ref{fig:bkg}). We consider a grid of normalization values for the ALP spectrum, $N_{\textnormal{tot}}$, between $8.4\times10^{-60}$~cm$^{-2}$ and $8.4 \times 10^{-50}$ cm$^{-2}$, motivated by \textit{Fermi}-LAT’s expected flux sensitivity. For each of the three background levels, we add the normalized ALP spectrum (60 steps within the range quoted above) and produce 2000 simulations for each data realization for two different progenitor masses (10 and 18-M$_\odot$), resulting in a total of 720,000 simulated spectra. Finally, we conduct spectral fitting for each spectrum, considering two different spectral models. The first model is the ALP model described in Section~\ref{sec:ALPmodel} with one free parameter, the total normalization $N_{\textnormal{tot}}$, on top of the background model described by a power law with the normalization and power-index as free parameters. The second model is the background-only fit. For both cases, we use \textit{XSPEC}'s \texttt{pgstat} statistical method, which describes Poisson data with Gaussian background \cite{GiacomoPGSTAT}. Finally, we utilize Wilks' theorem as applied to the scenario in which the ALP signal: given a large number of realizations of data, the test statistic $\Lambda= -2 \log \left(L_{\textnormal{null}}/L_{\textnormal{alternative}}\right)$, follows a half-$\chi^2$ distribution \cite{wilks1938, Cowan:2010js}, when no additional ALP signal is injected. In our case, $L_{\textnormal{null}}$ is the background-only fit, $L_{\textnormal{alternative}}$ is the ALP model fit, and the difference in the number of degrees of freedom (d.o.f) is one (ALP normalization). For each background consideration, we find a critical signal normalization for which we claim that the ALP model is preferred---taken to be when the log-likelihood analysis indicates half of the $\Lambda$ values that would have probabilities less than $5.7\times 10^{-7}$ if the background-only hypothesis was correct, corresponding to a 5$\sigma$ detection. We use 60 grid steps within the $N_\text{tot}$ normalization range quoted above, with 30 of them a refinement, to accurately determine this ``turn-over'' point. An example of $\Lambda$ distributions for a 10-M$_\odot$ progenitor for the median background (GRB 121011A) is shown in Fig.~\ref{fig:Wilks}.
\par
\begin{figure*}[]
\centering
\subfloat{\includegraphics[width=0.5\textwidth]{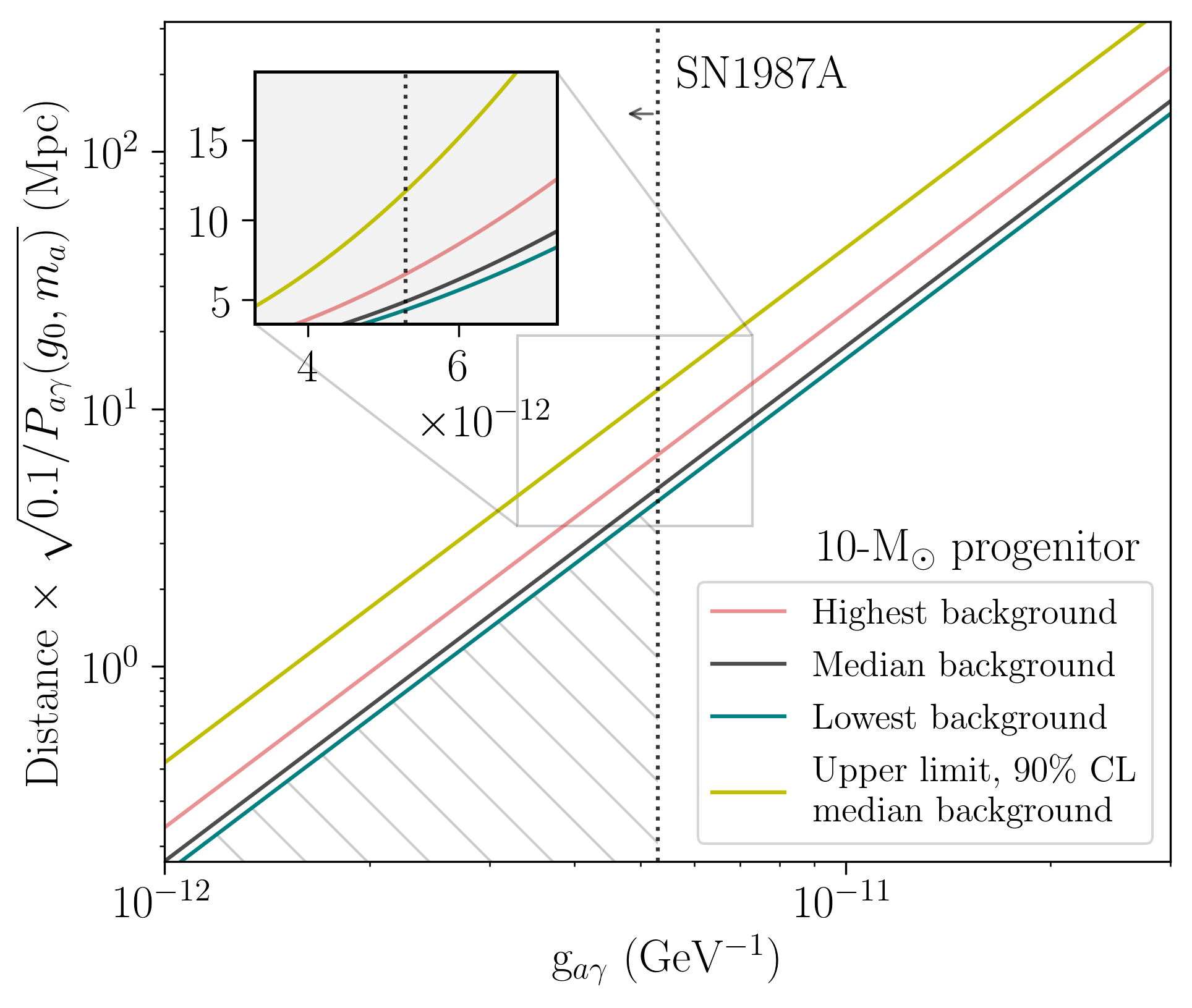}}
\subfloat{\includegraphics[width=0.5\textwidth]{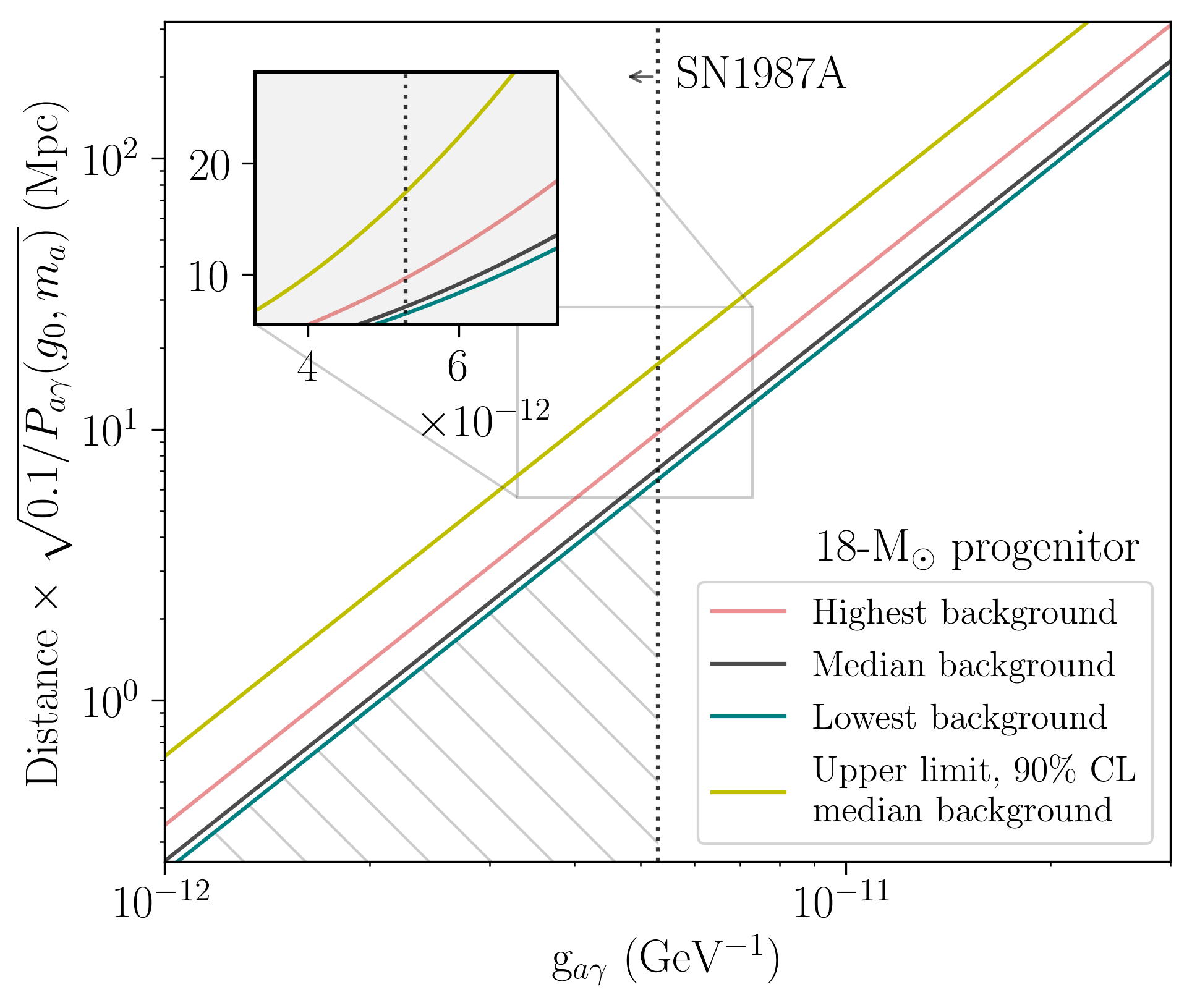}}
\caption{Distance limits for a LAT-LLE ALP detection for a 10-M$_\odot$ progenitor (left panel), and an 18-M$_\odot$ progenitor (right). The green, black, and red solid lines represent the 5-$\sigma$ detection limits on distances with so-far observed background levels from our GRB sample with \textit{Fermi} LLE, while the yellow solid line represents the expected $90\%$~CL upper limit for the median background. The dotted vertical line in each plot shows the upper limit for the ALP-photon coupling, $g_{a\gamma} = 5.3 \times 10^{-12}$ GeV$^{-1}$, derived in \cite{SN1987A_Payez}. The hatching in both plots shows the parameter space to which \textit{Fermi} LLE is sensitive, taking into consideration the SN1987A upper limit on the ALP-photon coupling.}
\label{fig:dist-coup}
\end{figure*}

The values of $\Lambda$ corresponding to normalization values of$N_{\textnormal{tot}} = 3.46 \times 10^{-54}$ cm$^{-2}$ for GRB 100826A (lowest background, high $\theta$), $N_{\textnormal{tot}} = 2.76 \times 10^{-54}$ cm$^{-2}$ for GRB 121011A (median background, medium $\theta$), and $N_{\textnormal{tot}} = 1.51 \times 10^{-54}$ cm$^{-2}$ for GRB081024A (highest background, low $\theta$), for a 10-M$_\odot$ progenitor, favor the ALP model over the background-only model. For a given ALP coupling $g_{a\gamma}$, Fig.~\ref{fig:dist-coup} shows the maximum allowed distance to a CCSNe for which a 5-$\sigma$ ALP signal discovery can be expected (provided the given gamma-ray background and $P_{a\gamma}$ assumed from Fig.~\ref{fig:conversion}). On the other hand, if a time- and distance-tagged CCSN is observed without any detected ALP signal, then the yellow curve gives, on average, the expected 90\;\%~CL upper limit to be derived on the ALP coupling $g_{a\gamma}$. The right panel of Fig.~\ref{fig:dist-coup} shows an analogous analysis, using the 18-M$_\odot$ progenitor. For the lowest background (high $\theta$), the corresponding normalization value is $N_{\textnormal{tot}} = 1.56 \times 10^{-54}$ cm$^{-2}$; for the median background (medium $\theta$) it is $N_{\textnormal{tot}} = 1.30 \times 10^{-54}$ cm$^{-2}$; and finally, for the highest background (low $\theta$) it is $N_{\textnormal{tot}} = 7.06 \times 10^{-55}$ cm$^{-2}$. The corresponding distance limits for the deduced upper bound on coupling from the SN1987A analysis, $g_{a\gamma} = 5.3 \times 10^{-12}$ GeV$^{-1}$ \cite{SN1987A_Payez}, in addition to a consideration of different conversion probabilities, are summarized in  Table~\ref{tab:results}. We remark that at the detection limit, only a few ALP-induced gamma-ray photons would be detected ($\sim$10 counts in the LLE sample), which makes it challenging to reliably reconstruct the energy spectrum or to alone trigger a GRB signal detection (considering the look-elsewhere effect in full sky surveys). Finally, we conclude the sensitivity analysis by noting that the distance limit variations in Fig.~\ref{fig:dist-coup} are driven by differences in LLE effective area---thus, the lower background count-rates, shown in Fig.~\ref{fig:bkg}, are mainly a consequence of decreasing detector acceptances at higher $\theta$s.

\begin{table}[]
\caption{Maximum distance to the ALP source to be within reach of the \textit{Fermi}-LLE sensitivity. We assume the ALP-photon coupling $g_{a\gamma} = 5.3 \times 10^{-12}$ GeV$^{-1}$ \cite{SN1987A_Payez} for different ALP-photon conversion probabilities, $P_{a\gamma}$, as seen in Fig.~\ref{fig:conversion}. Distance limits are in Mpc, shown for a 10-M$_\odot$ progenitor on the left, and 18-M$_\odot$ in parentheses on the right, for different background levels. Note that the different background levels are dependent on the LLE effective area, which decreases with an increase in $\theta$, hence lowering the event rate in the detector (see Fig.~\ref{fig:bkg}). This, however, is not an indicator of the intrinsic GRB background levels, but rather detected counts in the instrument.}
\begin{tabular}{clll}
\hline
\hline
\multicolumn{1}{l}{Conversion probability} & \multicolumn{3}{c}{ Distance limit (Mpc)}                                                                   \\
\multicolumn{1}{c}{$P_{\gamma}(g_0)$}    & \multicolumn{3}{c}{ \textit{Background level:}}  \\
                                        & \textit{Low}               & \textit{Median}            & \textit{High}              \\
\hline
 0.1                                        & 4.4 (6.5)         & 4.9 (7.1)        & 6.6 (9.7)          \\
0.05                                      & 3.1 (4.6)         &3.5 (5.0)       &4.7 (6.9)        \\
 0.01                                       &1.4 (2.1)          &  1.5 (2.3)         & 2.1 (3.1)          \\
 0.001                           &  0.4 (0.7)& 0.5 (0.7) &  0.7 (1.0) \\
\hline
\hline
\end{tabular}
\label{tab:results}
\end{table}

%% file: GRBs_analysis.tex

We consider the selected sample of unassociated GRBs in Table~\ref{tab:GRBlist} and conduct a spectral fitting for each GRB to find the highest significance for an inclusion of the ALP spectral component. Although it is unlikely that a nearby star's core-collapse would remain an undetected CCSN to the current optical all-sky surveys, such as \cite{Pojmanski:2002kn,Shappee:2013mna,2018PASP..130f4505T,Woniak_2004,Drake_2009,Law_2009}, or that an ordinary GRB would arrive without a time delay to the ALP signal from the core-collapse, we here make the ansatz to search for an ALP signal only within the detected GRB signal time window. We consider a null model to include components commonly used to describe ordinary GRB emission, and compare it to the alternative including the additional ALP component. 

\subsection{Data Preparation}
\label{subsec:data_prep}

The data in our sample was obtained from the public \textit{Fermi Science Support Center} (FSSC) website \footnote{\url{https://fermi.gsfc.nasa.gov/ssc/data/}, accessed on April 23, 2019.}. To analyze data, we use the \textit{Fermi} Science Tools \footnote{\url{https://fermi.gsfc.nasa.gov/ssc/data/analysis/software/}, accessed on April 23, 2019.} in combination with the HEAsoft \textit{XSPEC} spectral fitting software \footnote{\url{https://heasarc.gsfc.nasa.gov/lheasoft/}, accessed on April 23, 2019.}. We conduct a combined analysis between GBM, LLE, and standard LAT transient data using the analysis tools commonly used in the high-energy transient community \cite{XSPEC_code}. 
\par
The GBM analysis is done using \textit{gtburst} \footnote{\url{https://github.com/giacomov/gtburst}}. We conduct a binned analysis of the GBM data. Firstly, we compute the overall signal-to-noise ratio (SNR) for twelve GBM detectors and consider the three strongest signals recorded in the NaI detectors and one signal from a BGO detector for the spectral analysis. To determine the background, we use \textit{gtburst} to specify off-signal intervals, fit a polynomial to each channel of the detector, and interpolate these polynomials to compute a background spectrum over a given time interval. For each GBM detector, we consider the same time interval of the burst, determined by visual inspection, approximately corresponding to the flattening of the light curve with the background level. The corresponding GRB duration is listed as T$_{95}$ in Table~\ref{tab:GRBlist}, reflecting the time interval reported in Table III in \cite{Ajello_2019}. Finally, we produce spectral and background files appropriate for the analysis in \textit{XSPEC}.
\par
The preparation of the LLE data follows the same pathway 
as that of the GBM data. We assume the same burst  duration as determined by the GBM value of T$_{95}$ in Table~\ref{tab:GRBlist}. 
\par
With its lower number of counts, LAT transient data requires a different approach utilizing an unbinned analysis. Due to the transient nature of the source, we make use of the event class \texttt{P8R3\_TRANSIENT020\_v2}, analyzed with the corresponding galactic and isotropic diffuse templates \footnote{\url{https://fermi.gsfc.nasa.gov/ssc/data/access/lat/BackgroundModels.html}, as accessed on April 23, 2019.}, over a time interval determined by the T$_{95}$ values referenced in Table~\ref{tab:GRBlist}. Using \textit{gtburst}, for most GRBs, we perform a zenith cut of 100$\degree$, with a few-degree variation depending on a given source. We conduct a maximum likelihood analysis of the LAT source to obtain a LAT counts map within the considered region of interest (RoI, usually 12\degree). Once we obtain likelihood fit result parameters using \texttt{gtlike}, we proceed onto creating energy-binned background files using \texttt{gtbkg} and spectral files of energy-binned signal counts using \texttt{gtbin}, both readable by \textit{XSPEC}. Throughout this analysis, we use the point-source localization information provided in the \textit{Fermi} LAT Second GRB Catalog, 2FLGC \cite{Ajello_2019}.

\subsection{\textit{XSPEC} analysis}
\label{subsec:XSPEC}

We conduct a standard spectral fitting procedure of the selected GRB sample \cite{GRBAnalysis}. Fitting is conducted in \textit{PyXspec}, an object-oriented Python interface to \textit{XSPEC} \footnote{\url{https://heasarc.gsfc.nasa.gov/xanadu/xspec/python/html/index.html}, accessed on April 23, 2019.}. When modeling the spectral shape of a given GRB, we consider models commonly used in GRB spectral fitting, including single power law (denoted \texttt{pow} in Table~\ref{tab:GRBlist}), power law with a high-energy exponential cut-off (also known as ``comptonized model'', \texttt{cutoffpl}), the phenomenological Band function \cite{1993ApJ...413..281B} (\texttt{grbm}), or a combination thereof. We also include a consideration of an additional thermal component in the form of a blackbody spectrum (\texttt{bb}), as suggested in \cite{Guiriec:2010zj, GRBAnalysis}. Appendix~\ref{app:models} provides the details about the used models. We apply \textit{XSPEC}'s \texttt{pgstat} statistical method \cite{GiacomoPGSTAT} and find the fit with the lowest test statistic (in \textit{XSPEC} denoted by \texttt{PG-statistic}), obtaining profile log-likelihood (LL) values from the combined GBM, LLE and LAT transient data. We note that the ALP spectral model may be well-reproduced for a specific range of parameters of the Band function; in particular, for a 10-M$_{\odot}$ progenitor in Fig.~\ref{fig:ALPspec} the corresponding Band model parameters are $\alpha_1 \simeq 2.4$, $\alpha_2 \simeq -0.1,$ and $E_c \simeq 30$ MeV. However, in the considered GRB sample, these parameter values are not reached, as shown in Table~\ref{tab:GRBlist}, and are not expected for ordinary GRB spectral shapes. Similarly, the ALP spectral shape can be reproduced reasonably well with a blackbody function described by a peak temperature of $\sim 70$~MeV; however, all the GRBs listed in Table~\ref{tab:GRBlist} that are best fit by including a blackbody spectral component (GRBs 100826A, 110721A, 140102A, and 160917A) peak at the keV temperatures. For each model, we step through the neighboring fit parameter values to ensure that the best-fit parameters found from the maximum LL analysis are likely global, and not local minima.  
\par
To compare two nested models, we apply the log-likelihood ratio (LLR) test $\Lambda= -2 \log \left(L_{\textnormal{GRB}}/L_{\textnormal{GRB+ALP}}\right)$, with $L_{\textnormal{GRB}}$ corresponding to 
the likelihood of the null hypothesis, i.e.\ 
the GRB model constituting of commonly used functions for ordinary GRB emission. $L_{\textnormal{GRB+ALP}}$ corresponds to the alternative hypothesis, constituting of an ALP signal added on top of the null model, with all the considered parameters left free to vary. Table~\ref{tab:GRBlist} contains results of the model comparison of our data sample, using the 10-M$_{\odot}$-progenitor ALP spectral model. We note that the 18-M$_{\odot}$-progenitor has an almost identical spectral shape, leaving the results essentially independent on these two progenitors masses.

\subsection{ALPs from GRBs: fitting results}

\begin{figure}
    \centering
\includegraphics[width = .49\textwidth]{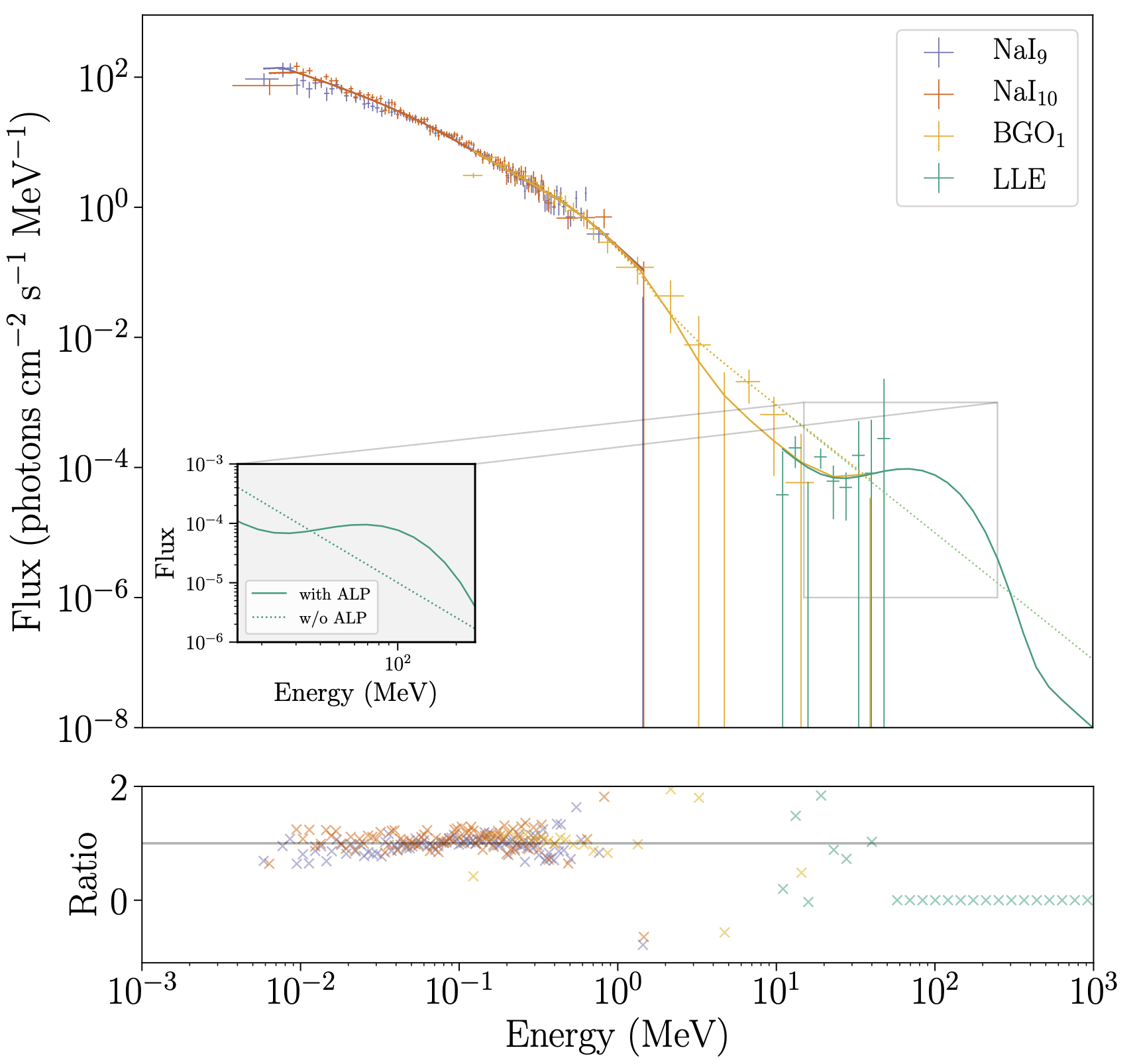}
\caption{GRB 101123A $\gamma$-ray flux with two different overlaid models. Different colors represent counts obtained by different detectors: purple, red, yellow, and blue correspond to NaI$_{9}$, NaI$_{10}$, and BGO$_{1}$ detectors on GBM respectively; and green corresponds to the LLE transient data. The solid lines represent unfolded model fits for each instrument when the additional ALP component is included, whereas the dotted lines are the best-fit model without the additional ALP component. The inplot shows the difference between the alternative and the null model in more detail: the solid green line represents the alternative model and the dotted line is the null model. The alternative model is composed of a Band function, exponential cut-off function, and the ALP signal. Finally, the lower panel shows the ratio of the observed data to the corresponding model that includes the ALP component, i.e.\ data/model. For plotting purposes only, the GBM data is shown binned, whilst LLE data is shown with the original binning provided by the instrument.}
\label{fig:model_inplot}
\end{figure}

None of the GRBs in the considered sample showed a significant improvement in the fit when including the additional ALP signal component. The model comparison for one GRB, 101123A, indicates a $\Lambda$ value of 5.8, corresponding to $\sim2.4 \sigma$ detection, pre-trials. The data and the fitted models for GRB 101123A are shown in Fig.~\ref{fig:model_inplot} with an inplot showing the difference between the best null model and the alternative model with the additional ALP component. 

{}

This alternative best-fit hypothesis has an ALP component with the normalization $N_\text{tot}= 4.9\times10^{-52}$ cm$^{-2}$. Applying the coupling $g_{a\gamma}=5.3\times10^{-12}$ GeV$^{-1}$ and a conversion probability $P_{a\gamma} = 0.01$ (see Fig.~\ref{fig:conversion} for this GRB's sky position), we find that the corresponding distance would be $\sim 120$~kpc.
Note that GRB~101123A was observed strongly off-axis, at an incidence angle of $\sim$~80$\degree$. Under such conditions, LLE's effective area decreases significantly. Thus, even a source with such a large $N_\text{tot}$ value of the ALP component results in only few counts and no reconstructed energies above $\sim$50~MeV.
\par
We then include the trials factor \cite{Lyons:1900zz}, to take into account the size of the considered parameter space, and express the global significance by $p_\text{global} = 1-(1-p_{\textnormal{local}})^{\textnormal{N}_{\textnormal{trials}}}$. From the local $p$-value, $p_{\textnormal{local}} = 1.6 \times 10^{-4}$ and the number of GRB trials in our sample, $N_{\textnormal{trials}}=24$, this results is a global $p$-value of $\sim0.3$, further indicating that this observation is not statistically significant.

%% file: conclusion.tex
In this paper, we consider the light ALPs produced via the Primakoff process in a collapse of a massive star which, by converting into photons in the Galactic magnetic field, could produce an observable gamma-ray flux. The duration of an ALP burst is expected to be on the order of $\sim$10 seconds. Due to its uncertain and likely negligible effect, we do not take into consideration the ALP-photon conversion that may take place in the magnetic field of the intergalactic medium and, due to a lack of magnetic field models, we do not take into consideration ALP-photon conversions that may take place within the host galaxy. In fact, the contribution from the host galaxy would increase the observed gamma-ray flux, rendering our current results conservative. Furthermore, due to the complexity of core-collapse modeling, we only consider two CCSN progenitor masses: 10 and 18 M$_\odot$ \cite{SN1987A_Payez}. However, theoretical considerations suggest that long GRBs are produced in explosions of very massive ($\gtrsim 20$ M$_\odot$, \cite{Levan:2018}) stars which, in turn, would produce a higher number of ALPs as compared to a lower mass progenitor. Thus, the combination of our magnetic field and progenitor mass choices renders our reported results conservative.
\par
We find the sensitivity of the \textit{Fermi}-LAT instrument using the LLE data sample including energies $\gtrsim 30$~MeV to detect ALP-induced gamma-ray emission from \mbox{CCSNe} for ALP masses \mbox{$m_a \lesssim 10^{-10}$} eV. In particular, we consider a sample of GRB backgrounds and compute the maximum allowed distance to core-collapsing stars that still give statistically significant, 5$\sigma$, ALP-signal detection. For the lowest background,
we obtain that the limiting distance is $\sim$3~Mpc for the conversion probability, $P_{a\gamma} = 0.05$, for a 10-M$_{\odot}$ progenitor, and $\sim$5~Mpc for an 18-M$_{\odot}$ progenitor, for a coupling of $g_{a\gamma} = 5.3 \times 10^{-12}$ GeV$^{-1}$ \cite{SN1987A_Payez}. For the highest background count, the farthest distance corresponds to $\sim$5~Mpc and $\sim$7~Mpc; and for the median background, it is $\sim$3.5~Mpc and $\sim$5~Mpc for 10- and 18-M$_\odot$ progenitors respectively. Finally, the distance limits reported in this paper, in addition to the observed background levels, are driven by LLE effective area variation that tends to decrease for observations at larger instrumental incidence angles---and thus lower background count rates.
These limiting distances for an ALP-signal detection from a CCSN
for different conversion probabilities and axion-photon couplings $g_{a\gamma}$ are shown in Table ~\ref{tab:results} and Fig~\ref{fig:dist-coup}. 
The results found in this paper by utilizing the LLE data cut technique and its resulting data sample are comparable to those done with the standard LAT analysis in \cite{axionscope}. 
As such, conducting a search for an ALP signal from a close-by CCSN using the LLE technique, independent from or in parallel with the standard LAT analysis, can be a useful way of probing the ALP parameter space. Furthermore, the distance limits found in this investigation may be complemented by utilizing the upper energy range of the better-resolved GBM data, or the rest of the LAT transient data, to search for the tail distribution from the ALP-induced gamma-ray emission. 
\par
Finally, we consider a sample of unassociated, thus potentially nearby, LLE-detected GRBs (see Table ~\ref{tab:GRBlist}). We conduct a spectral model fitting for each candidate using the \textit{XSPEC} library models commonly used for GRB spectral modeling. Once the best-fit for an ordinary GRB spectrum is determined, we conduct an analogous modelling procedure by introducing an additional ALP spectral component. We find that all of the GRB emissions in our sample are well-fitted by commonly used GRB spectral models and that introducing an additional ALP spectral component does not result in a statistically significant improvement. 
\par
In this paper, we assume that the ALP-induced gamma-ray signal itself triggers the GRB observation or that the ALP signal from a CCSN coincides  with the ordinary GRB signal, which is unlikely to be the case. 
The main source of uncertainty is determining the core-collapse time, and thus the expected arrival time of any ALP-induced GRB. Therefore, an interesting investigation would be a dedicated search for potential ALP induced gamma-ray photons arriving before the GRB trigger times. 
As suggested in \cite{optical}, using optical lightcurves to predict explosion times may be another way of attempting these analyses, as done in \cite{Meyer:2020vzy}. Nevertheless, the optimal resolution for the time-tagging issue is using neutrino detection from a CCSN, followed by a search for ALP emission in the coincident gamma-ray observation; although, at this time, no such coincident detection has been confirmed in association with a GRB. 
\par
The GRB model comparison analysis does not allow for a deduction of the limits on the ALP coupling, $g_{a\gamma}$. In order to obtain such information, we would require the GRBs' distance information, as done in \cite{Meyer:2020vzy}. With the current and upcoming optical surveys such as ASAS-SN \cite{ASAS-SN}, ZTF \cite{ZTF}, TESS \cite{TESS}, and the Vera C. Rubin Observatory \cite{LSST}, the number of the observed nearby CCSNe (e.g. $z<0.02$ or 100 Mpc), \cite{Meyer:2020vzy}) is likely to increase. This, in turn, would improve the probability of \textit{Fermi} detecting their corresponding GRBs, allowing for a statistically significant study of such sources in the context of ALP searches and limits on the relevant ALP coupling space. An example of such an analysis determining the upper limits on the ALP coupling is shown in \cite{Meyer:2020vzy}.
\par
Furthermore, with the new generation of gamma-ray instruments, such as e-ASTROGAM \cite{e-ASTROGAM}, ComPair \cite{compair}, PANGU \cite{pangu}, or alike, the improved sensitivity and angular resolution particularly at energies relevant to the ALP signal ($<$100 MeV) and FoV's similar to \textit{Fermi} LAT, the search for ALP-induced GRBs will be substantially improved. In particular, observatories such as AMEGO \cite{AMEGO}, with its excellent sensitivity, angular and energy resolution, low energy threshold, and a large field of view, will allow for the most stringent constraints on the ALP parameter space, surpassing the limits of the current ALP laboratory experiments \footnote{\url{https://www.snowmass21.org/docs/files/summaries/CF/SNOWMASS21-CF3_CF2_Regina_Caputo-122.pdf}, as accessed on April 6, 2021.}. 
\par
Finally, besides CCSNe, additional astrophysical objects may be considered as sites of ALP production. In particular, a production of ALPs has been hypothesized during neutron-star (NS) mergers; albeit further theoretical work is needed to constrain the expected ALP spectrum from such events \cite{Harris:2020qim}. Taking into consideration the most recent observations from NS mergers using e.g. LIGO/Virgo \cite{GW}, as well as a rapid development of the field of gravitational wave astronomy, using gravitational waves may be yet another probe into the production time and nature of ALPs in the future.

%% file: acknowledge.tex
The authors thank Miguel A. Sánchez-Conde for discussion and detailed feedback. M.~C. acknowledges support by NASA under award number 80GSFC21M0002. M.~M. acknowledges support from the European Research Council (ERC) under the European Union’s Horizon 2020 research and innovation program Grant agreement No. 948689 (AxionDM) and from the Deutsche Forschungsgemeinschaft (DFG, German Research Foundation) under Germany’s Excellence Strategy -- EXC 2121 ``Quantum Universe’’ -- 390833306.

The \textit{Fermi} LAT Collaboration acknowledges generous ongoing support
from a number of agencies and institutes that have supported both the
development and the operation of the LAT as well as scientific data analysis.
These include the National Aeronautics and Space Administration and the
Department of Energy in the United States, the Commissariat \`a l'Energie Atomique
and the Centre National de la Recherche Scientifique / Institut National de Physique
Nucl\'eaire et de Physique des Particules in France, the Agenzia Spaziale Italiana
and the Istituto Nazionale di Fisica Nucleare in Italy, the Ministry of Education,
Culture, Sports, Science and Technology (MEXT), High Energy Accelerator Research
Organization (KEK) and Japan Aerospace Exploration Agency (JAXA) in Japan, and
the K.~A.~Wallenberg Foundation, the Swedish Research Council and the
Swedish National Space Board in Sweden.
 
Additional support for science analysis during the operations phase is gratefully
acknowledged from the Istituto Nazionale di Astrofisica in Italy and the Centre
National d'\'Etudes Spatiales in France. This work performed in part under DOE
Contract DE-AC02-76SF00515.
\newline

%% file: appendix.tex

\section{{GRB Models}}
\label{app:models}
To fit the selected GRB sample, we use \textit{XSPEC} models that include:
\begin{enumerate}
    \item Band function (\texttt{grbm}, gamma-ray burst continuum), described by
    \begin{equation}
    A(E) = 
    \begin{cases}
    K E^{\alpha_1} \exp(-E/E_c), \quad \text{if} \quad E < E_c(\alpha_1 - \alpha_2)\\
    K [(\alpha_1-\alpha_2)E_c]^{(\alpha_1-\alpha_2)} \exp(\alpha_2 - \alpha_1) E^{\alpha_2},\\
    \hspace{3.5cm} \text{otherwise}
\end{cases}
    \end{equation}
    where $E$ is the energy in units of keV. Model parameters are $\alpha_1$, first power law index; $\alpha_2$, second power law index; $E_c$, characteristic energy in keV; and $K$ is the normalization constant in units of photons/keV/cm$^2$/s.

    \item Power law, (\texttt{pow}), described by:
    \begin{equation}
        A(E) = KE^{-\alpha},
    \end{equation}
    where $\alpha$ is the power-law index.
    
    \item Power law with  high energy exponential cut-off, (\texttt{cutoffpl}), described by
    \begin{equation}
        A(E) = KE^{-\alpha}\textnormal{exp}(-E/\beta),
    \end{equation}
    where $\beta$ is the e-folding energy of the exponential rolloff (in keV). 
    \item Blackbody spectrum, (\texttt{bb}), described by
     \begin{equation}
        A(E) = K \frac{E^2}{\textnormal{exp}(E/kT)-1},
    \end{equation}
    where $kT$ is the temperature in keV.
\end{enumerate}